\documentclass[amsmath,amssymb,aps,twocolumn,superscriptaddress,prl,10pt]{revtex4-2}

\usepackage{graphicx, braket}
\graphicspath{{figures/}}
\usepackage[normalem]{ulem}
\usepackage{float}
\usepackage{hyperref}

\DeclareMathOperator{\Tr}{Tr}

\DeclareMathOperator{\const}{const}
\DeclareMathOperator{\sign}{sign}

\renewcommand{\Re}{\operatorname{Re}}

\begin{document}

\title{Measurement-induced L{\'e}vy flights of quantum information}

\author{Igor Poboiko}
\thanks{These authors contributed equally to this work.}
\affiliation{\mbox{Institute for Quantum Materials and Technologies, Karlsruhe Institute of Technology, 76131 Karlsruhe, Germany}}
\affiliation{\mbox{Institut f\"ur Theorie der Kondensierten Materie, Karlsruhe Institute of Technology, 76131 Karlsruhe, Germany}}

\author{Marcin Szyniszewski}
\thanks{These authors contributed equally to this work.}
\affiliation{\mbox{Department of Physics and Astronomy, University College London, Gower Street, London, WC1E 6BT, UK}}
\affiliation{\mbox{Department of Computer Science, University of Oxford, 15 Parks Rd, Oxford OX1 3QD, United Kingdom}}

\author{Christopher J. Turner}
\affiliation{\mbox{Department of Physics and Astronomy, University College London, Gower Street, London, WC1E 6BT, UK}}

\author{Igor V. Gornyi}
\affiliation{\mbox{Institute for Quantum Materials and Technologies, Karlsruhe Institute of Technology, 76131 Karlsruhe, Germany}}
\affiliation{\mbox{Institut f\"ur Theorie der Kondensierten Materie, Karlsruhe Institute of Technology, 76131 Karlsruhe, Germany}}

\author{Alexander D. Mirlin}
\affiliation{\mbox{Institute for Quantum Materials and Technologies, Karlsruhe Institute of Technology, 76131 Karlsruhe, Germany}}
\affiliation{\mbox{Institut f\"ur Theorie der Kondensierten Materie, Karlsruhe Institute of Technology, 76131 Karlsruhe, Germany}}

\author{Arijeet Pal}
\affiliation{\mbox{Department of Physics and Astronomy, University College London, Gower Street, London, WC1E 6BT, UK}}

\date{\today}

\begin{abstract}

We explore a model of free fermions in one dimension,
subject to \emph{frustrated} (non-commuting) local measurements across adjacent sites, which resolves the fermions into non-orthogonal orbitals, misaligned from the underlying lattice.
For maximal misalignment, superdiffusive behavior emerges from the vanishing of the measurement-induced quasiparticle decay rate at one point in the Brillouin zone, which generates fractal-scaling entanglement entropy $S \propto \ell^{1/3}$ for a subsystem of length $\ell$.
We derive an effective non-linear sigma model with long-range couplings responsible for L{\'e}vy flights in entanglement propagation, which we confirm with large-scale numerical simulations.
When the misalignment is reduced, 
the entanglement exhibits, with increasing $\ell$, consecutive regimes of superdiffusive, $S\propto \ell^{1/3}$, diffusive, $S\propto \ln \ell$, and localized, $S = \rm{const}$, behavior.
Our findings show how intricate fractal-scaling entanglement can be produced for local Hamiltonians and measurements.
\end{abstract}

\maketitle

\textit{\textbf{Introduction.}}---
Quantum dynamics in many-body systems subjected to measurements has attracted much attention. It was, in particular, shown that quantum measurements may induce transitions between phases with the different scaling of entanglement entropy $S$ as a function of subsystem size $\ell$ \cite{Li2018a, Skinner2019a, Chan2019a, Cao2019a, Szyniszewski2019a,  Li2019a, Bao2020a} (see also reviews on monitored quantum circuits~\cite{Potter2022, Fisher2022}). A special role in this context is played by systems of free complex fermions with local density measurements preserving the Gaussian character of the state \cite{Cao2019a, Alberton2021a, Buchhold2021a, Coppola2022, Carollo2022, 
Szyniszewski2022, Poboiko2023a, Poboiko2023b, chahine2023entanglement, Lumia2023, starchl2024generalized, FavaNahum2024}. It was shown that in one-dimensional (1D) geometry, $S(\ell)$ saturates as $\ell \to \infty$ (area law) \cite{Poboiko2023a, FavaNahum2024} (cf.\@ Refs.~\cite{Cao2019a, Coppola2022}). For small measurement rate $\gamma$, it is preceded by an intermediate range of $\ell$ with the scaling  $S\sim \gamma^{-1} \ln \ell$. In ${d>1}$ dimensions, a measurement-induced transition between an area-law phase and a phase with $\ell^{d-1} \ln \ell$ scaling of $S$ is found \cite{Poboiko2023b, chahine2023entanglement}. 
There is a remarkable relation between the physics of monitored systems in $d$ dimensions and Anderson localization in disordered systems in $d+1$ dimensions, with the area law for $S(\ell)$ corresponding to the localized phase and the $S \propto \ell^{d-1} \ln \ell$ behavior to the diffusive phase (or diffusive regime for $d=1$). This relation can be inferred from the comparison of the respective field theories---non-linear sigma models (NLSMs)---for the two problems \cite{Jian2023, Fava2023, Poboiko2023a, Poboiko2023b, chahine2023entanglement, starchl2024generalized, FavaNahum2024, guo2024, Poboiko2025, tiutiakina2024}.

Importantly, local 
measurements on free fermions prevent establishing a volume-law phase (${S \propto \ell^{d}}$) \cite{Fidkowski2021}, 
which is a typical phase in generic weakly-monitored quantum circuits. The appearance of the volume-law phase for fermions requires interactions between particles~\cite{guo2024, Poboiko2025}, which breaks down the Gaussianity of the many-body states.
Even more tricky is to obtain a \emph{fractal} sub-extensive scaling of entanglement ($S\propto \ell^{\zeta}$ with ${d-1<\zeta<d}$) in monitored systems, although it was reported for, e.g., space-time dual quantum circuits~\cite{Ippoliti2022}, long-range interacting Hamiltonians or unitary gates~\cite{Block2022a, Sharma2022, Richter2023}, or ``long-range dissipation and monitoring'' \cite{deAlbornoz_LRLindblad, Russomanno2023}. In this context, non-commutativity (\textit{``frustration''}) of measurement operators (among themselves or with respect to the unitary dynamics of the system \cite{Bao2020a, Lunt2020a, VanRegemortel2021a, Fava2023, Lumia2023, Richter2023,Nehra2025}) is expected to be of crucial importance for the phase diagrams. Another interesting class of models is based on the \textit{measurement-only} dynamics~\cite{Ippoliti2021a,Klocke2023,Nehra2025}, where the quasi-local, possibly non-commuting measurements give rise to both generation and suppression of entanglement. 

\begin{figure}[bt]
    \centering
    \includegraphics[width=\columnwidth]{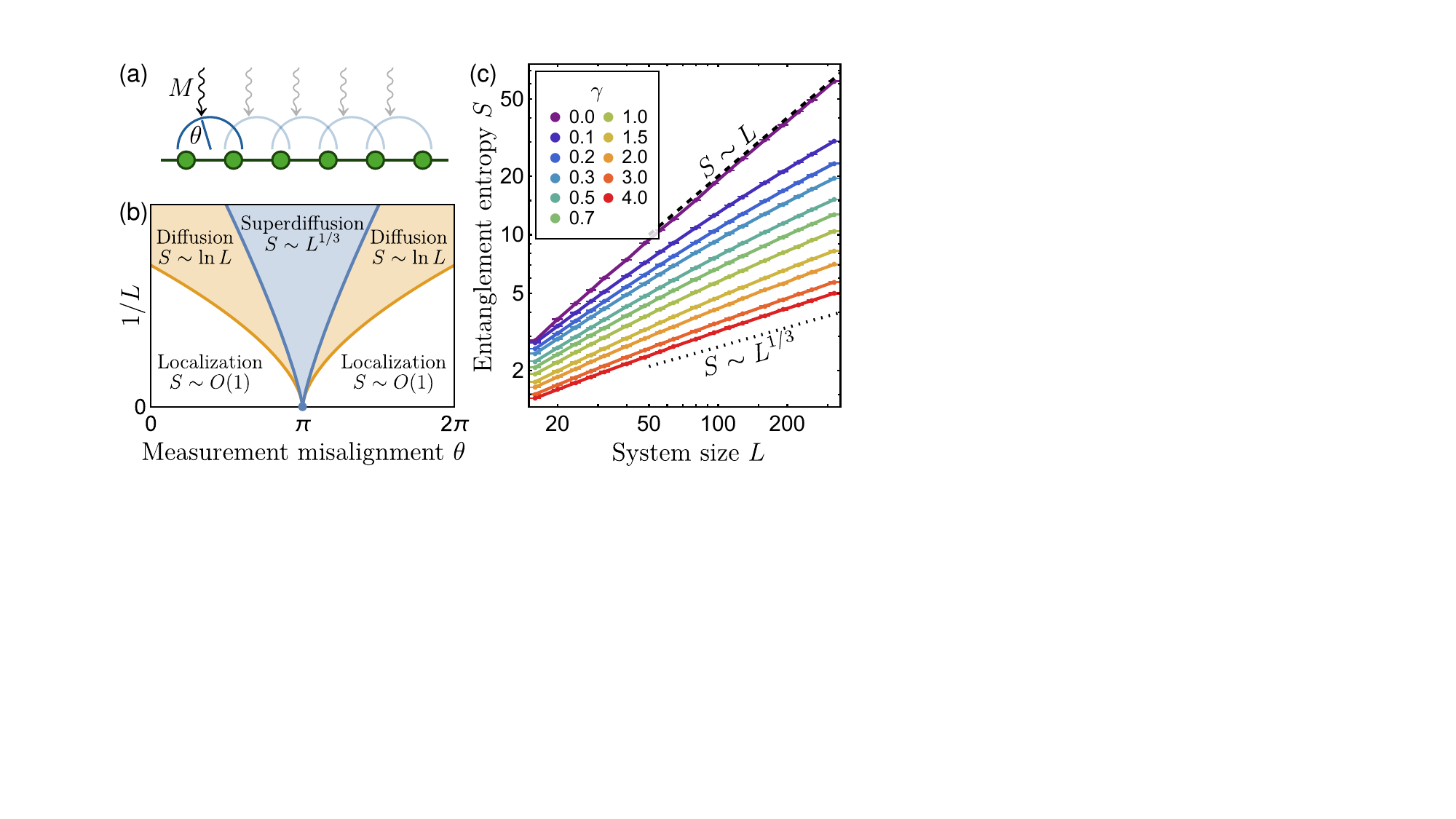}
    \caption{(a)~Model: free-fermion chain of size $L$ continuously monitored via frustrated measurements with a misalignment $\theta$. (b)~``Phase diagram'' of this system showing superdiffusion at $\theta=\pi$, with characteristic regimes of the entanglement scaling indicated by different colors. (c)~Half-chain entanglement entropy $S$ as a function of $L$ for different values of the measurement strength $\gamma$ at fixed $\theta = \pi$. The dashed line shows the extensive (``ballistic'') behavior $S\sim L$, while the dotted line shows the fractal scaling $S\sim L^{1/3}$ corresponding to superdiffusion in 1+1-dimensional space-time. 
    }
    \label{fig:entropy}
\end{figure}

In this Letter, we explore a 1D model of monitored free fermions, with the measurement operator being a particle number in a state residing on two adjacent sites (rather than on a single site), Fig.~\ref{fig:entropy}. This extension affects neither the $\mathrm{U}(1)$ symmetry (particle-number conservation) nor the local character of the measurement operator, nor the Gaussianity of the states. In view of the universality of diffusion and localization for given symmetry class and spatial dimensions, one could thus expect that the model belongs to the same ``universality class'' as previously studied 1D fermionic models~\cite{Poboiko2023a}. Remarkably, this is not always the case. We discover the emergence of the physics of superdiffusion (L{\'e}vy flights) of quantum information, 
with a fractal power-law scaling $S(\ell) \propto \ell^{1/3}$, which persists into the measurement-only limit. We emphasize that, at variance with previous works, such behavior is observed in an intrinsically \textit{short-ranged free-fermion} model.

\textit{\textbf{Model.}}---We consider a model of monitored free fermions with non-commuting measurements and U(1) particle-number symmetry, on a periodic chain of $L$ sites [Fig.~\ref{fig:entropy}(a)].
The monitored dynamics is characterized by the stochastic Schr\"odinger equation~\cite{Cao2019a, Wiseman2009},
\begin{align}
\label{eq:SSE}
  d\ket{\psi_{t+dt}} &= \Big[-i dt H - \frac{\gamma\, dt}{2} \sum_i (M_i - \langle M_i \rangle )^2 \nonumber \\
  & \quad + \sum_i d\xi_i^t (M_i - \langle M_i \rangle) \Big] \ket{\psi_t},
\end{align}
where $\gamma$ is the measurement strength, and $d\xi^t_i$ is the It\^o increment with variance $\gamma\, dt$.
The Hamiltonian includes fermion hopping, $H = J\sum_i c^\dagger_{i+1} c_i + \text{h.c.}$, where $J$ is the hopping strength.
This stochastic evolution describes continuous monitoring of $M_i$~\cite{Wiseman2009}, which we define to be the following two-site operator
\begin{equation}
    M_i = d_i^\dagger d_i, \quad d_i = c_i \cos \frac{\theta}{4} + c_{i+1} \sin \frac{\theta}{4},
    \label{eq:measurement-operator}
\end{equation}
where parameter $\theta$ can be interpreted as a misalignment of the measurement apparatus that causes a superposition of two adjacent sites to be measured. When this misalignment is nonzero, the measurement operators on adjacent bonds do not commute, leading to frustration in the chain, with the maximal frustration happening for $\theta = \pi$. 

This continuous monitoring can be realized through weak interaction of every pair of adjacent sites with its ancilla \cite{Turkeshi2021, Doggen2022a, Doggen2023}; the latter is then projectively measured with a period $dt$. Note that the specific sequence of measurements within the time interval $dt$ becomes immaterial in the continuous-time limit $dt\to 0$.
Since this evolution preserves the Gaussianity of the state, it is computationally simulable in polynomial time, and any state properties can be calculated from the corresponding single-particle correlation matrix ${\cal G}_{ij} = \braket{c_i^\dagger c_j}$. In detail, we use the algorithm of Refs.~\cite{Cao2019a, Alberton2021a, Szyniszewski2022, Szyniszewski2024}, where the state is represented as an $N\times L$ matrix ($N=L/2$ being the number of particles), and the evolution involves multiplication by $L\times L$ matrices corresponding to $H$ and $M_i$, see Supplemental Material (SM)~\cite{SuppMat}.

Judging from the previous analytical description of monitored free fermions with the U(1) symmetry~\cite{Poboiko2023a}, one would be tempted to conclude that at a large enough spatial scale, this model should exhibit localization (i.e., area-law entanglement), with an intermediate logarithmic regime at small $\gamma$. However, a numerical analysis of the entanglement entropy at $\theta=\pi$ in Fig.~\ref{fig:entropy}(c) (where $\ell = L/2$) indicates the absence of localization, even at large measurement strengths. Furthermore, the data surprisingly reveal an entanglement growth that is faster than logarithmic. As we demonstrate analytically later, and support by a thorough numerical analysis, the entropy grows as $S \propto \ell^{1/3}$, which corresponds to a superdiffusive transport in 1+1 (space-time) dimensions. 

Note that measurement operators similar to Eq.~\eqref{eq:measurement-operator} were employed in a model of monitored Majorana fermions~\cite{Kells2023, Jian2023, Fava2023, Merritt2023, Nehra2025}, which violates the $\mathrm{U}(1)$ symmetry: there, a superposition of Majorana operators at adjacent sites was considered. The physics in these works is related to logarithmic anti-localization quantum corrections in symmetry classes D and DIII. This is very different from L{\'e}vy-flight-induced superdiffusion of quantum information leading to the power-law fractal scaling of entanglement entropy studied here.

\textit{\textbf{Effective field theory.}}---
Our analytical description of the problem is based on the replicated Keldysh NLSM approach, developed in Refs.~\cite{Poboiko2023a, Poboiko2023b, Poboiko2025} and extended to weak measurements in Ref.~\cite{chahine2023entanglement}, see SM~\cite{SuppMat} for details. We introduce the Keldysh fermionic path-integral representation defined on $R \to 1$ replicas of the Keldysh contour. Averaging over the white noise $\xi(t)$ present in Eq.~\eqref{eq:SSE} leads to the quartic fermionic term in the action. This term is decoupled by means of the Hubbard-Stratonovich matrix-valued field $\hat{Q}(x,t)$, which is interpreted as the local equal-time Green's function of $d$-fermions, $Q_{\alpha\beta}(x,t)\sim2\langle d_{\alpha}(x,t)d_{\beta}^{\ast}(x,t)\rangle $. Here, indices include the structure in the Keldysh and replica spaces, $\alpha, \beta \in \{+, -\}_{\text{K}} \otimes \{1, \dots, R\}_{\text{R}}$, and the spatial coordinate $x$ is a continuous version of the lattice index $i$. The Goldstone manifold consists of a replica-symmetric sector, the two-dimensional sphere $\mathrm{S}^2$, which describes the Lindbladian dynamics, and a replicon sector, the special unitary group $\hat{U} \in \mathrm{SU}(R)$, which describes the dynamics of observables that are non-linear in the density matrix. 

The crucial observation, which is responsible for the superdiffusive spreading of the quantum information in the system, is that the measurement-induced quasiparticle decay rate
has the following momentum-dependent form:
\begin{equation}
\label{eq:gammak}
\gamma_{k}=\gamma\left(1+\sin\frac{\theta}{2}\,\cos k\right),
\end{equation}
and vanishes at $k = \pm\pi$ for a special point $\theta = \pi$. It happens because, for $\theta = \pi$, the operator $d_i = (c_i + c_{i+1}) / \sqrt{2}$ (and hence $M_i$) exactly nullifies the state with $k = \pm \pi$. Thus, such states are completely unaffected by measurements. Interestingly, this does not affect the diffusive behavior of the Lindbladian dynamics observed earlier \cite{Poboiko2023a} in the conventional density monitoring case $\theta = 0$. The spatial diffusion coefficient consists of two contributions attributed to the unitary dynamics and non-commutativity of measurements,
\begin{multline}
\label{eq:DiffusionCoefficient}
D=\int_{-\pi}^{\pi}\frac{(dk)}{\gamma_{k}}\left[\left(\frac{\partial \xi_{k}}{\partial k}\right)^{2}+\frac{1}{4}\left(\frac{\partial \gamma_{k}}{\partial k}\right)^{2}\right]\\
=\frac{4J^{2}}{\gamma}\frac{1}{1+\left|\cos\frac{\theta}{2}\right|}+\frac{\gamma}{4}\left(1-\left|\cos\frac{\theta}{2}\right|\right),
\end{multline}
and remains finite at $\theta = \pi$, since both the group velocity $\xi^\prime_k = - 2 J \sin k$ and the derivative $\gamma^\prime_k$  vanish at $k = \pm\pi$ similar to $\gamma_k$. 

We now focus on the replicon sector, which describes observables of our interest.
To see the emergence of superdiffusion, we inspect the quadratic form of the action (see SM~\cite{SuppMat} for the full action of NLSM) in $1+1$ dimensions.
In the spatial direction, it has a conventional diffusive form characterized by the diffusion coefficient $D$, Eq. (\ref{eq:DiffusionCoefficient}). On the other hand, vanishing of $\gamma_k$ leads to non-locality of the temporal term in the action
characterized by the diffusion kernel ${\cal B}(t_1-t_2)$,
with the Fourier transform 
given by
\begin{multline}
{\cal B}(\omega)=\int_{-\pi}^{\pi}\frac{(dk)}{\gamma_{k}-i\omega}=\frac{1}{\sqrt{(\gamma-i\omega)^{2}-\gamma^{2}\sin^{2}\frac{\theta}{2}}}\\
\stackrel{\omega\to 0}{\approx} \begin{cases}
\left(\gamma\left|\cos\frac{\theta}{2}\right|\right)^{-1}, & \theta\neq\pi \,,\\
\left[-2i\gamma(\omega+i0)\right]^{-1/2}, & \theta=\pi \,.
\end{cases}
\label{eq:B-omega}
\end{multline}
The emergence of superdiffusive L{\'e}vy flights with exponent $\alpha=3/2$---resulting in a heavy-tailed distribution of quantum-information spreading---in our 
theory at $\theta=\pi$ is manifest in the last line of Eq.~\eqref{eq:B-omega}.
The physics behind superdiffusion at $\theta=\pi$ is as follows.
The states with $k$ close to $\pm \pi$ are nearly eigenstates of measurement operators $M_i$ 
and thus propagate ballistically for long times $\sim 1/\gamma_k$ [see Eq.~\eqref{eq:gammak}] before they get substantially affected by measurements that limit quantum correlations.

The superdiffusive character of the field theory leads to the fractal scaling of observables. We focus below on the entanglement entropy (for a subsystem A of length $\ell$) 
$S_{A} = -\overline{\Tr\left(\hat{\rho}_{A}\ln\hat{\rho}_{A}\right)}$ and the charge correlation function 
\begin{equation}
    C(x-x^{\prime})=\overline{\left\langle \hat{n}(x)\hat{n}(x^{\prime})\right\rangle -\left\langle \hat{n}(x)\right\rangle \left\langle \hat{n}(x^{\prime})\right\rangle },
    \label{Cx}
\end{equation} 
where the overbar denotes averaging over quantum trajectories. Let us emphasize that $C(x-x^\prime)$ is a fundamental characteristic of the system, which governs the scaling of various key observables. In particular, it determines the second cumulant of charge ${\cal C}_{A}^{(2)}$,
\begin{equation}
\label{eq:C2}
{\cal C}_{A}^{(2)}
=\overline{\left\langle \hat{N}_{A}^{2}\right\rangle -\left\langle \hat{N}_{A}\right\rangle ^{2}}
=\int_{0}^{\ell}\!dx\,dx^{\prime}\,C(x-x^{\prime}),
\end{equation}
which, in view of the Gaussian character of the state, is related to the entropy via:
\begin{equation}
S_{A}\approx (\pi^{2}/3) \: {\cal C}_{A}^{(2)}.
\label{eq:ratio-S-C2}
\end{equation}
The exact relation \cite{KlichLevitov} also contains terms proportional to higher cumulants. However, they do not affect the scaling and amount to a small correction only, as was found for conventional density monitoring \cite{Poboiko2023a}; we have also verified this for the present model \cite{SuppMat}. Both the entropy and the charge-cumulant generating function can be expressed via the NLSM partition function with appropriate boundary conditions \cite{Poboiko2025, SuppMat}.

\begin{figure}
    \centering
    \includegraphics[width=\columnwidth]{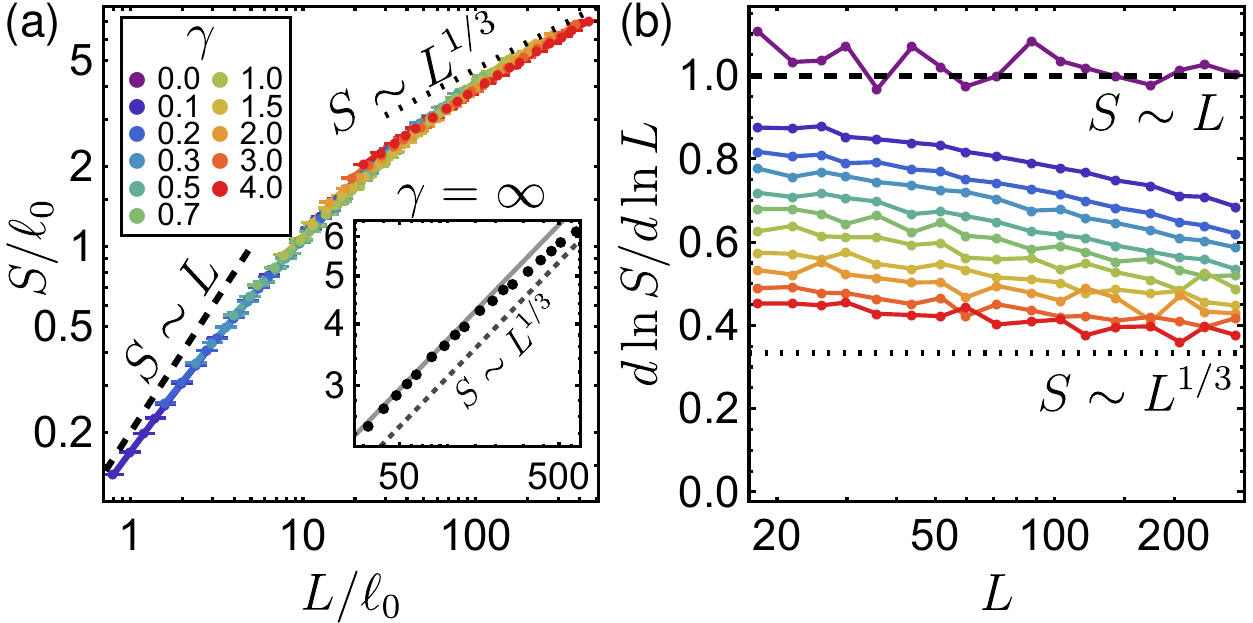}
    \caption{(a)~Collapse of $S / \ell_0$, where $S$ is half-chain entanglement entropy, as a function of $L / \ell_0$ for $\theta = \pi$ and measurement rates from $\gamma=0$ to $\gamma=4$. The inset shows results for $\gamma=\infty$, which are of the form $S = s(L) L^{1/3}$, where $s(L)$ exhibits a slow crossover from a finite value at $L \sim 30 \ell_0$ to a slightly smaller finite value at $L \to \infty$ due to quantum corrections, see SM \cite{SuppMat}.
    (b) $d \ln S / d \ln L$ as a function of the system size. Legend in (a) applies in (b). The dashed lines show the ballistic behavior $S\sim L$, while the dotted lines show superdiffusion $S\sim L^{1/3}$ in the thermodynamic limit.}
    \label{fig:ent_collapse}
\end{figure}

\textit{\textbf{Fractality of correlations and entanglement.}}---
We first consider the problem within the quasiclassical approximation with the Gaussian action.
For $\delta = \theta - \pi \ll 1$, the Fourier transform of Eq.~\eqref{Cx} reads
\begin{equation}
\label{eq:C-q}
  C(q) \approx  \begin{cases}
    (2|\delta|)^{-1/2} |q \ell_0|,        & \,q \ell_0 \ll |\delta|^{3/2}, \\
    (2^{-2/3} 3^{-1/2}) |q \ell_0|^{2/3}, & \,|\delta|^{3/2} \ll q \ell_0 \ll 1,\\
    \const, & \, 1 \lesssim q \ell_0,
  \end{cases}
\end{equation}
where 
$\ell_0 = \sqrt{D / \gamma} \approx \sqrt{(2J/\gamma)^2 + 1/4}$ is the mean free path.
Thus, $\delta = 0$ ($\theta = \pi$) is a critical point, where $C(q) \propto q^{2/3}$ for $q \ell_0 \to 0$, and the system exhibits a fractal (superdiffusive) scaling of the charge cumulant and entropy, 
\begin{equation}
\label{eq:S-scaling}
S_{A} \sim {\cal C}_{A}^{(2)} \sim \ell_0 (L/\ell_0)^{1/3} \quad \text{for}\ \ \ell= L/2\, \gg \ell_0\,,
\end{equation}
explaining the surprising numerical results from Fig.~\ref{fig:entropy}. At small measurement strength $\gamma$, the entropy for small system sizes scales extensively with the system size $S \sim L$, which then experiences a ballistic-to-superdiffusion crossover, corresponding to the crossover between the second and third lines of Eq.~\eqref{eq:C-q}, as one increases $L$ or $\gamma$. This change in the entanglement behavior can be seen in Fig.~\ref{fig:ent_collapse}(a), where we observe a nearly perfect data collapse of $S / \ell_0$ vs $L / \ell_0$ in a broad range of $\gamma$, from 0.1 to 4.0. Analytically, this universality of the crossover function is, strictly speaking, derived for $\gamma \ll 1$, in view of numerical corrections to the prefactor of $L^{1/3}$ scaling at $\gamma \gtrsim 1$, which come from spatial scales of the order of the lattice spacing and are not included in the NLSM analysis. We see, however, from Fig.~\ref{fig:ent_collapse}(a) that the universality holds excellently up to a large measurement rate, $\gamma = 4$. This universality shows that quantum corrections are essentially irrelevant even for rather large $\gamma$. A similar theory, with diffusive transport along one axis and superdiffusive along the other axis, was derived for transport in graphene with anisotropic disorder \cite{Gattenloehner16Levy}. It was found there that quantum localization amounts to a finite correction only, without inducing strong localization or a localization transition.  The ballistic-to-superdiffusion crossover is further demonstrated in Fig.~\ref{fig:ent_collapse}(b), where the logarithmic derivative $d\ln S / d\ln L$ is shown to approach the 1/3 asymptotic value in the large-$L$ limit, $L / \ell_0 \to \infty$. Our conclusions are additionally supported by an analysis of finite-size corrections to the superdiffusive scaling in SM~\cite{SuppMat}.

Remarkably, the superdiffusive behavior and the absence of localization also hold in the measurement-only case, $\gamma = \infty$, 
see inset of Fig.~\ref{fig:ent_collapse}(a),
in agreement with the analytical results. At the same time, the $\gamma = \infty$ data
slightly deviate (bends down) from the universal scaling curve. This has two reasons. First, for $\gamma = \infty$, quantum corrections to the prefactor of the $L^{1/3}$ scaling mentioned in the preceding paragraph are particularly pronounced. Second, the $\gamma = \infty$ model belongs to a different symmetry class, as we are going to explain. 
A free-fermion system with particle-number conservation belongs to the BDI symmetry class when the Hamiltonian exhibits a particle-hole symmetry $H = -H^T$ \textit{and}, in the same basis, the measurement operators are real $M = M^*$; otherwise, it belongs to the AIII class \cite{FavaNahum2024, Poboiko2025}. For any finite $\gamma$, our model is therefore in the AIII class, while, for $\gamma = \infty$, the Hamiltonian is absent in the stochastic Schr\"odinger equation and the measurement-only point has a larger BDI symmetry. The difference between the NLSM field theories of these two classes is minimal and does not affect the qualitative behavior. 
The one-loop weak-localization correction in class BDI is negative and twice larger than that for class AIII. In our model, this is expected to lead to a numerical reduction of the prefactor $s$ of $S= s L^{1/3}$ scaling in the $\gamma = \infty$ case (BDI class). This is what is observed in our simulations (inset of Fig.~\ref{fig:ent_collapse}(a)): $s(L)$ slowly interpolates between two finite values, as expected from the weak-localization correction, see SM \cite{SuppMat}.

\begin{figure}
    \centering
    \includegraphics[width=\columnwidth]{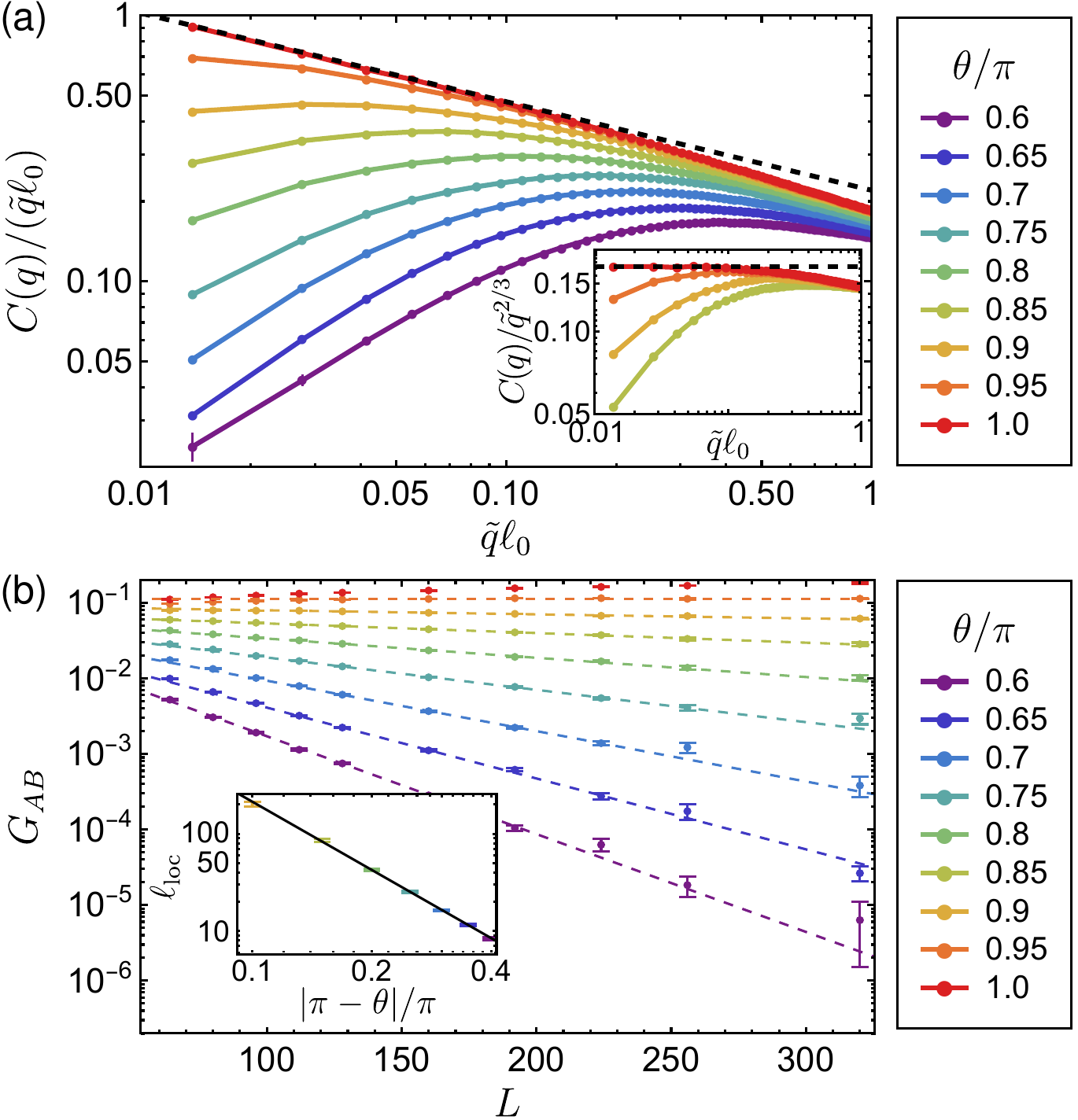}
    \caption{(a)~Correlation function $C(q)$ as a function of rescaled momentum $\tilde q \ell_0$,  where $\tilde q = 2 \sin(q/2)$ and $\ell_0 = \sqrt{(2 J / \gamma)^2 +1/4}$, for $\gamma=4$ and different values of $\theta/\pi$. The system size used is $L = 320$. The dashed line shows the superdiffusive behavior $C(q)/q \sim q^{-1/3}$. The inset shows $C(q) / q^{2/3}$, which saturates at $q\to 0$ for $\theta=\pi$.  (b)~Particle number covariance $G_{AB}$ as a function of system size $L$ for different values of $\theta/\pi$. Dashed lines are fits to $\sim\nobreak\exp(-L/4\ell_\text{loc})$ for $L\ge 128$. The extracted localization length is shown in the inset, along with a power-law fit $\ell_\text{loc} \sim |1-\theta/\pi|^\nu$, with exponent $\nu\approx 2.33(3)$ consistent with the analytical asymptotics \eqref{eq:localization_length} at numerically accessible scales.}
    \label{fig:Cq}
\end{figure}

For a non-zero (but small) $\delta$, the system exhibits a crossover from superdiffusive regime $C(q)\propto q^{2/3}$ to diffusive regime $C(q)\propto q$ at momentum $q \ell_0 \sim |\delta|^{3/2}$, which corresponds to a lengthscale $\ell^* \sim \ell_0 |\delta|^{-3/2}$.
Ultimately, at large system sizes, the system then crosses over into localization, $C(q)/q\to 0$.
We confirm this behavior using finite-size numerics in Fig.~\ref{fig:Cq}(a), where we plot the ratio $C(q)/q$. The three distinct regimes are clearly observed: superdiffusion (dashed line), diffusion (approximate saturation, with a slow decrease towards small $q$ due to weak-localization correction), and localization [vanishing $C(q)/q \propto q$ at $q \to 0$].
Translating to the real space, this implies that, as the system size is increased, one will first see the fractal (superdiffusive) entropy scaling  $S \propto L^{1/3}$, then the logarithmic (diffusive) law $S \propto \ln L$, and finally the area law (localization) $S \simeq \text{const}$, see Fig.~\ref{fig:entropy}(b).

In the diffusive regime, one can calculate the effective coupling constant (discarding localization effects),
\begin{equation}
\left.   
g =  \frac{C(q)}{q} \right|_{\displaystyle q\ell^* \ll 1} 
  \simeq \frac{\ell_0}{\sqrt{2}} |\delta|^{-1/2}.
  \label{eq:conductance}
\end{equation}
This allows us to estimate the localization length, which scales at $|\delta| \ll 1$ as
\begin{equation}
  \ell_\text{loc} \sim \ell^* \exp(4\pi g) \simeq \ell_0 |\delta|^{-3/2} \exp \frac{2\sqrt{2} \pi \ell_0}{\sqrt{|\delta|}}.
  \label{eq:localization_length}
\end{equation}
The localization length $\ell_\text{loc}(\delta)$ thus diverges exponentially at the critical point $\delta=0$. If one fits Eq.~\eqref{eq:localization_length} to a power-law form $\ell_\text{loc} \sim |\delta|^{-\nu}$ in a restricted range of $\delta$, one will get an ``effective exponent'' $\nu$ that increases from 3/2 towards infinity when one approaches the singular point $\delta=0$. 

To probe the localization in real space numerically, we use the particle-number covariance, $G_{AB} = \overline{\langle N_A \rangle \langle N_B \rangle - \langle N_A N_B \rangle}$, where $N_A$ and $N_B$ are particle-number operators of antipodal regions $A$ and $B$, each of size $L/4$. This observable yields an effective conductance at scale $\sim L/4$ and scales in the area-law phase asymptotically 
 as $G_{AB} \sim \exp(- L / 4\ell_\text{loc})$. Numerical results for $G_{AB}(L)$ shown in Fig.~\ref{fig:Cq}(b) confirm that localization sets in when $\theta \ne \pi$. By fitting $G_{AB}(L)$ to the exponential law for $L \ge 128$ (dashed lines), we obtain estimates for the localization length $\ell_{\rm loc}(\delta)$ shown in the inset. 
 The results clearly support the analytically predicted divergence of $\ell_{\rm loc}$ at $\delta\to 0$. Accurately verifying  Eq.~\eqref{eq:localization_length} in this way is hardly possible since $\ell_{\rm loc}$ quickly becomes very large at small $\delta$, where this formula holds. Instead, we show in the inset a power-law fit $\ell_\text{loc} \sim |\delta|^{-\nu}$, which yields effective exponent $\nu \approx 2.33(3)$, which is larger than 3/2 in agreement with a discussion below Eq.~\eqref{eq:localization_length}. 
 Note that this effective exponent is not too far from 3/2, which reflects the difficulty in numerical verification of the exponential dependence in Eq.~\eqref{eq:localization_length}: when the exponential decay of $G_{AB}(L)$ is observed for realistic system sizes, the localization length is only a few times larger than $\ell^* \sim \ell_0 |\delta|^{-3/2}$. 

\textit{\textbf{Discussion and outlook.}}---
Summarizing, the free-fermion model with two-site monitoring operators  \eqref{eq:measurement-operator} is characterized, at $\theta=\pi$, by a superdiffusive NLSM, which leads to the fractal scaling of the entanglement entropy, $S \propto L^{1/3}$, and charge correlations, $C(q)\propto q^{2/3}$. When $\theta$ deviates from the critical point $\pi$, the superdiffusive scaling is transient, giving rise to diffusion and, eventually, localization at longer length scales (smaller $q$). The superdiffusive behavior originates from the vanishing of the measurement-induced quasiparticle decay rate $\gamma_k$ at one point in the Brillouin zone and should also hold for other models of monitoring having this property (also for other symmetry classes).   An example is a model with measurement of conventional site density $c_i^\dagger c_i$ but on even sites only. 
A related mechanism of L{\'e}vy flights was found to be operative in non-monitored models with ``nodal'' disorder~\cite{Gattenloehner16Levy, Wang2024}, dephasing~\cite{Wang2023}, and interactions~\cite{Wang2025}.
Our discovery of measurement-induced L\'evy flights opens up avenues for the study of exotic entangled phases and novel quantum-information transport mechanisms, including tailoring the entanglement growth in free-fermion monitored systems.

\textit{\textbf{Acknowledgments.}}---
A.~P.\@ and M.~S.\@ were funded by the European Research Council (ERC) under the European Union's Horizon 2020 research and innovation programme (Grant Agreement No.\@ 853368). M.~S.\@ was also supported by the Engineering and Physical Sciences Research Council grant on Robust and Reliable Quantum Computing (RoaRQ), Investigation 004 [grant reference EP/W032635/1]. C.~J.~T.\@ is supported by an EPSRC fellowship (Grant Ref.\@ EP/W005743/1). I.~P. and A.~D.~M. acknowledge support by the Deutsche Forschungsgemeinschaft (DFG, German Research Foundation) -- 553096561. The authors acknowledge the use of the UCL High Performance Computing Facilities (Myriad), and associated support services, in the completion of this work.

\textit{\textbf{Data availability.}}--- The data that support the findings
of this article are openly available \cite{Data}.

\bibliography{refs}

\vspace{1cm}

\widetext
\clearpage

\setcounter{secnumdepth}{1}
\renewcommand{\theequation}{S\arabic{equation}}
\renewcommand{\thefigure}{S\arabic{figure}}
\renewcommand{\thesection}{S\arabic{section}}
\setcounter{equation}{0}
\setcounter{figure}{0}

\begin{center}
\Large \textbf{Supplemental Material to ``Measurement-induced L{\'e}vy flights of quantum information''}
\end{center}

\section{Derivation of superdiffusive NLSM}

Here, we derive the effective action of the NLSM field theory for the measurement protocol considered in the present Letter. The derivation is based on the approach developed in Refs.~\cite{Poboiko2023a, Poboiko2023b, Poboiko2025}. The crucial difference is that now we monitor the density of $d$-fermions, which are linearly related to the original fermions $c$. An additional difference is that the present protocol corresponds to continuous monitoring, as opposed to projective measurements studied in Refs.~\cite{Poboiko2023a, Poboiko2023b}; this, however, does not qualitatively affect the results. 

\subsection{Fermionic action for weak measurements}

Continuous measurements of an arbitrary operator $\hat{M}$ can be realized via a complete set of Kraus operators enumerated with an index $\alpha \in \mathbb{R}$ corresponding to possible measurement outcomes:
\begin{equation}
\label{eq:analytics:Krauss}
\hat{\mathbb{K}}_{\alpha}=\exp\left[-\gamma dt\left(\hat{M}-\alpha\right)^{2}\right],\qquad \sqrt{\frac{2\gamma dt}{\pi}}\int_{-\infty}^{+\infty} d\alpha \, \hat{\mathbb{K}}_{\alpha}^{\dagger}\hat{\mathbb{K}}_{\alpha}=\hat{\mathbb{I}}
\end{equation}
Following Ref.~\cite{Poboiko2025}, we introduce a replicated fermionic Keldysh path integral representation, and first focus on each individual replica and a given ``measurement trajectory''---that is, a realization of measurement outcomes $\{\alpha_m\}$, where $m$ enumerates discrete time steps. The Lagrangian consists of a replica-diagonal contribution describing the unitary evolution:
\begin{equation}
{\cal L}_{0}[c]=c^{\dagger}\left(i\partial_{t}-\hat{H}_{0}\right)c,\qquad c\equiv\begin{pmatrix}c_{+}\\
c_{-}
\end{pmatrix}_{\text{K}},\qquad c^{\dagger}\equiv\begin{pmatrix}c_{+}^{\ast} & -c_{-}^{\ast}\end{pmatrix}_{\text{K}},
\end{equation}
and a measurement-induced contribution that acquires the following form:
\begin{equation}
\prod_{m}\mathbb{K}_{\alpha_{m}}^{(+)}(t_{m})\mathbb{K}_{\alpha_{m}}^{(-)}(t_{m})\simeq
\exp\left[i\int dt\,{\cal L}_{M}(\xi(t))\right],\qquad
i{\cal L}_{M}(\xi)=-\gamma \sum_{s=\pm}\left(\frac{\xi(t)}{2\gamma}-M_{s}(t)+\frac{1}{2}\right)^{2},
\end{equation}
where continuous (in the limit $dt \to 0$) field $\xi(t_{m}) \equiv \gamma (2\alpha_{m}-1)$ is introduced.

For the case of density monitoring, $\hat{M} = \hat{d}^\dagger \hat{d}$, where $\hat{d}$ is an arbitrary (different from $\hat{c}$) set of fermions, we employ the ``principal value'' regularization procedure described in detail in Ref.~\cite{Poboiko2023a} and arrive at the ``symmetrized'' coherent state representation $M(d^\ast, d) = 1/2 + d^\ast d$. Utilizing the Grassmann nature of fields $d$, we finally arrive at the following Lagrangian:
\begin{equation}
i{\cal L}_{M}(\xi,d)=-\frac{\xi^{2}}{2\gamma}+\xi\,d^{\dagger}\hat{\tau}_{z}d,
\end{equation}
where $\hat{\tau}_\alpha$ are the Pauli matrices acting in the Keldysh space. 

The full replicated action for the problem considered in the present Letter is then easily obtained by (i) introducing additional index $i$ enumerating the measurement operators (i.e. lattice sites), and performing summation over them, and (ii) performing summation over $R$ replicas, with the replica limit $R \to 1$ following from the Born's rule. 
Thus, the effect of weak measurements of fermionic density is described, within this replicated theory, by introducing a Gaussian white-noise field $\xi_i(t)$ with the correlation function $\left\langle \xi_{i}(t)\xi_{j}(t^{\prime})\right\rangle =(\gamma / R)\delta_{ij}\delta(t-t^{\prime})$ and coupling this noise to $d^\dagger \hat{\tau}_z d$ on the Keldysh contour. Finally, Gaussian integration over $\xi_i(t)$ yields a quartic ``interaction'', so that the full Lagrangian reads:
\begin{equation}
i{\cal L}=ic^{\dagger}\left(i \partial_t-\hat{H}_{0}\right)c+\frac{\gamma}{2R}\left(d^{\dagger}\hat{\tau}_{z}d\right)^{2} \,.
\end{equation}
Here, the (time- and space-dependent) fields $c$ and $d$ ``live'' in the $R$-dimensional replica space and, in addition, in the two-dimensional Keldysh space.

\subsection{Effective matrix field theory}

We perform the Hubbard-Stratonovich transformation to decouple the quartic term simultaneously in two possible channels, following the procedure from Ref.~\cite{Poboiko2023a}. First, we introduce an auxiliary matrix field $\hat{{\cal G}}$ via the functional delta-function $\delta_\text{slow}[\mathcal{F}]$, which, however, fixes only the ``slow'' components (introducing a momentum cutoff). We represent this delta-function through an auxiliary integral by introducing matrix field $\hat{\tau}_z \hat{Q}$:
\begin{equation}
1=\int{\cal D}{\cal G}\, \delta_{\text{slow}}\!\left[{\cal G}-dd^{\dagger}\tau_{z}\right]=\int{\cal D}{\cal G}{\cal D}Q\, \exp\left[-\frac{\gamma}{2R}\Tr\left(\hat{{\cal G}}\hat{\tau}_{z}\hat{Q}\right)-\frac{\gamma}{2R}d^{\dagger}\hat{Q}d\right].
\end{equation}
Next, we decouple the quartic term in two channels, utilizing the delta function; such a procedure is equivalent to applying Wick's theorem to the corresponding term (cf. Ref.~\cite{Poboiko2023a}):
\begin{equation}
\left(d^{\dagger}\hat{\tau}_{z}d\right)^{2}\simeq-\Tr\hat{{\cal G}}^{2}+\Tr^{2}\hat{{\cal G}}.
\end{equation}
We then perform Gaussian integration over $\hat{{\cal G}}$, arriving at:
\begin{equation}
\int{\cal D}{\cal G}\,\exp\left\{-\frac{\gamma}{2R}\left[\Tr\left(\hat{{\cal G}}^{2}+\hat{{\cal G}}\hat{\tau}_{z}\hat{Q}\right)-\Tr^{2}\hat{{\cal G}}\right]\right\}=\exp\left\{\frac{\gamma}{8R}\left[\Tr(\hat{\tau}_{z}\hat{Q})^{2}-\frac{1}{2R-1}\Tr^{2}(\hat{\tau}_{z}\hat{Q}_{0})\right]\right\}.
\end{equation}

As the final step, we note that for the problem we consider in this Letter, fermions $d$ and $c$ obey a linear relation with an auxiliary matrix $\hat{m}$
\begin{equation}
d=\hat{m}c,\quad m_{i,i}=\cos\frac{\theta}{4},\quad m_{i,i+1}=\sin\frac{\theta}{4}.
\label{matrix-m}
\end{equation}
Utilizing this relation and performing Gaussian integration over fermions $c$, we finally arrive at the following effective action for the ``slow'' matrix field $\hat{Q}$:
\begin{equation}
\label{eq:MatrixAction}
-S[\hat{Q}]=\Tr\ln\left(i \partial_t-\hat{H}_{0}+\frac{i \gamma}{2R}\hat{m}^{\dagger}\hat{Q}\hat{m}\right)+\frac{\gamma}{8R}\left\{\Tr\left[\left(\tau_{z}\hat{Q}\right)^{2}\right]-\frac{1}{2R-1}\Tr^{2}\left(\hat{\tau}_{z}\hat{Q}\right)\right\}.
\end{equation}

\subsection{Self-consistent Born approximation and saddle-point manifold}

First, we put $R = 1$ and focus on the replica-symmetric sector. As in earlier works, we note that for arbitrary $2 \times 2$ matrix $\hat{Q}_0$, there is an exact identity:
\begin{equation}
\Tr\left[\left(\hat{\tau}_{z}\hat{Q}_0\right)^{2}\right]-\Tr^{2}\left(\hat{\tau}_{z}\hat{Q}_0\right)\equiv\Tr^{2}\hat{Q}_0-\Tr\left(\hat{Q}_0^{2}\right),
\end{equation}
which implies that arbitrary unitary rotations $\hat{Q}_{0}\mapsto\hat{{\cal R}}\hat{Q}_{0}\hat{{\cal R}}^{-1}$ form a symmetry group of the action given by Eq.~\eqref{eq:MatrixAction}. The saddle-point equation (for a traceless matrix $\Tr \hat{Q}_0 = 0$) then yields:
\begin{equation}
\hat{Q}_0(\boldsymbol{r})=2i\left(\hat{m}\hat{G}_{Q_0}\hat{m}^{\dagger}\right)_{\boldsymbol{r}\boldsymbol{r}},\quad\text{where}\quad\hat{G}_{Q}=\left(i\partial_{t}-\hat{H}_{0}+i\frac{\gamma}{2}m^{\dagger}\hat{Q}\hat{m}\right)^{-1},
\label{Q0GQ}
\end{equation}
where $\boldsymbol{r} = (x,t)$ is the space-time coordinate.
When we look for a homogeneous solution,  $\hat{Q}_0(\boldsymbol{r}) = \hat{Q}_0 = \const$, this equation reduces to
\begin{equation}
\hat{Q}_{0}=2i\int\frac{(d\varepsilon)(dk)|m_{k}|^{2}}{\varepsilon-\xi_{k}+i\gamma_k \hat{Q}_{0} / 2}
\qquad \Longrightarrow \qquad
\hat{Q}_{0}
=\sign\hat{Q}_{0} \,,
\end{equation}
where we have introduced Fourier transform $m_k$ of matrix $\hat{m}$ and momentum-dependent decay rate $\gamma_k$ [given by Eq.~\eqref{eq:gammak} of the main text]:
\begin{equation}
m_{k}=\cos\frac{\theta}{4}+e^{ik}\sin\frac{\theta}{4},\quad\gamma_{k}\equiv\gamma|m_{k}|^{2}=\gamma\left(1+\sin\frac{\theta}{2}\,\cos k\right).
\end{equation}

Thus, an arbitrary matrix satisfying $\hat{Q}_0^2 = \hat{\mathbb{I}}$ and $\Tr \hat{Q}_0 = 0$ provides a solution. 
This yields the saddle-point manifold of the replica-symmetric sector of the theory, which is the two-dimensional sphere $\mathrm{S}^2$.
On this manifold, there is a special point (SCBA solution)
\begin{equation}
\hat{Q}_{\text{SCBA}}=\hat{\Lambda}=\begin{pmatrix}1-2n_{0} & 2n_{0}\\
2(1-n_{0}) & -(1-2n_{0})
\end{pmatrix}_{\text{K}},
\end{equation}
which yields the average value of the Green function, as determined by causality. Here
 $n_0$ is the filling factor of the band, which is conserved and thus determined by initial conditions. (Formally, $n_{0}=\left\langle \hat{d}_{i}^{\dagger}\hat{d}_{i}\right\rangle _{0}$, but average density for fermions $\hat{d}$ is same as for fermions $\hat{c}$.)
 The replica-symmetric manifold $\mathrm{S}^2$ is obtained by rotations of $\Lambda$ in Keldysh space.
 The full saddle-point manifold in the $R \to 1$ limit is then obtained by noting that, for $R \neq 1$, arbitrary rotations that commute with $\hat{\tau}_z$ also produce a symmetry, yielding the following parametrization:
 \begin{equation}
\hat{Q}=\begin{pmatrix}Q_{++}\hat{\mathbb{I}} & Q_{+-}\hat{U}\\
Q_{-+}\hat{U}^{\dagger} & Q_{--}\hat{\mathbb{I}}
\end{pmatrix}_{\text{K}},
\label{eq:QSPManifold}
\end{equation}
where $\hat{U} \in \mathrm{SU}(R)$.
 
\subsection{Gradient expansion}

We substitute $\hat{Q}=\hat{{\cal R}}\hat{\Lambda}\hat{{\cal R}}^{-1}$ in
the action \eqref{eq:MatrixAction} and identically rewrite the trace-log term  in a form suitable for gradient expansion:
\begin{equation}
-S[\hat{Q}]=\Tr\ln\left(1+\hat{G}_{\Lambda}\hat{{\cal R}}^{-1}\left[i \partial_t-\hat{H}_{0},\hat{{\cal R}}\right]+\frac{i\gamma}{2}\hat{G}_{\Lambda}\left(\hat{{\cal R}}^{-1}\hat{m}^{\dagger}\hat{{\cal R}}\hat{\Lambda}\hat{{\cal R}}^{-1}\hat{m}\hat{{\cal R}}-\hat{m}^{\dagger}\hat{\Lambda}\hat{m}\right)\right),
\label{eq:S-Q-tr-log}
\end{equation}
where $\hat{G}_{\Lambda}$ is given by $\hat{G}_{Q}$ from Eq.~(\ref{Q0GQ}) with $\hat{Q}\to \hat{\Lambda}$, matrices $\hat{m}$ are introduced in Eq.~(\ref{matrix-m}), and the limit $R \to 1$ was taken.

The first order of the expansion of Eq.~\eqref{eq:S-Q-tr-log} yields two terms. The first term is the standard Wess-Zumino term, contributing to the replica-symmetric sector only and governing its time dynamics:
\begin{equation}
\label{eq:app:S11}
-S_{t}^{(1)}[\hat{Q}]=\frac{1}{2}\Tr\left(\hat{\Lambda}\hat{{\cal R}}^{-1}\partial_{t}\hat{{\cal R}}\right).
\end{equation}
The second term arises due to the non-commutativity of measurements and gives a contribution to the diffusion coefficient:
\begin{equation}
-S_{x}^{(1)}[\hat{Q}]=\frac{\gamma}{8}\Tr\left(\left[\hat{m}^{\dagger},\hat{Q}\right]\left[\hat{m},\hat{Q}\right]\right)\approx-\frac{D_{1}}{8}\Tr\left(\partial_{x}\hat{Q}\right)^{2},\quad D_{1}=-\gamma\Tr\left(\left[\hat{m}^{\dagger},\hat{x}\right]\left[\hat{m},\hat{x}\right]\right).
\end{equation}
Yet another contribution to the diffusion coefficient arises from the second-order expansion in a standard way, and yields:
\begin{multline}
-S_{x}^{(2)}[\hat{Q}]=-\frac{1}{4}\Tr\Bigg[\hat{G}_{R}\left(1+\hat{\Lambda}\right)\hat{{\cal R}}^{-1}\partial_{x}\hat{{\cal R}}\left(\left[\hat{H}_{0},\hat{x}\right]-\frac{i\gamma}{2}\left(\hat{m}^{\dagger}\left[\hat{m},\hat{x}\right]-\left[\hat{m}^{\dagger},\hat{x}\right]\hat{m}\right)\right)\\
\times\hat{G}_{A}(1-\hat{\Lambda})\hat{{\cal R}}^{-1}\partial_{x}\hat{{\cal R}}\left(\left[\hat{H}_{0},\hat{x}\right] +\frac{i\gamma}{2}\left(\hat{m}^{\dagger}\left[\hat{m},\hat{x}\right]-\left[\hat{m}^{\dagger},\hat{x}\right]\hat{m}\right)\right)\Bigg]\approx-\frac{D_{2}}{8}\Tr(\partial_{x}\hat{Q})^{2},
\end{multline}
where 
\begin{equation}
D_{2}=\int(dk)\left\{\frac{1}{\gamma_{k}}\left[\left(\partial_k\xi_{k}\right)^{2}+\frac{1}{4}\left(\partial_k\gamma_{k}\right)^{2}\right]-\gamma|\partial_k m_{k}|^{2}\right\},
\end{equation}
and 
\begin{equation}
G_{R/A}(\varepsilon,k)=\frac{1}{\varepsilon-\xi_{k}\pm i\gamma_{k}/2}.
\end{equation}
The full diffusion coefficient $D = D_1 + D_2$ is then given by Eq.~\eqref{eq:DiffusionCoefficient} of the main text. 
Due to the non-locality of matrix $\hat{m}$, the diffusion coefficient $D$ is nonzero even in the measurement-only limit $J = 0$.

Finally, we focus on the term that governs the time dynamics of the replicon sector and is responsible for the L{\'e}vy flights. This term arises from the expansion beyond the first order with respect to the temporal-gradient term. It is sufficient to neglect the spatial dependence of $\hat{Q}$-matrix and focus on the temporal dependence only. Furthermore, we neglect fluctuations of the replica-symmetric sector here and make the following substitution:
\begin{equation}
\hat{{\cal R}}=\begin{pmatrix}\hat{{\cal V}}_{+} & 0\\
0 & \hat{{\cal V}}_{-}
\end{pmatrix}\ \Rightarrow\ \hat{{\cal R}}^{-1}i\partial_{t}\hat{{\cal R}}=\begin{pmatrix}i\hat{{\cal V}}_{+}^{-1}\partial_{t}\hat{{\cal V}}_{+} & 0\\
0 & i\hat{{\cal V}}_{-}^{-1}\partial_{t}\hat{{\cal V}}_{-}
\end{pmatrix},
\end{equation}
with ${\cal V}_{\pm} \in \mathrm{SU}(R)$ and $\hat{U} = \hat{{\cal V}}_{+} \hat{{\cal V}}_{-}^\dagger$.
The corresponding contribution to the action then reads:
\begin{equation}
S_{t}[Q]\approx -\int dx \int(dk)\Tr\ln\left(1+\hat{G}_{\Lambda}(k)\hat{{\cal R}}^{-1}(x,t)i\partial_{t}\hat{{\cal R}}(x, t)\right).
\end{equation}
Expanding this term to the second order, we obtain:
\begin{equation}
\label{eq:app:St2}
S_{t}^{(2)}[\hat{Q}]=n_{0}(1-n_{0})\int dxdt_{1}dt_{2}{\cal B}(t_{1}-t_{2})\Tr\left[\hat{{\cal J}}_t(x,t_{1})\hat{{\cal J}}_t(x,t_{2})\right],
\end{equation}
where the Fourier transform of the diffuson ladder block $\mathcal{B}(t_1 - t_2)$ is defined as
\begin{equation}
{\cal B}(\omega)=\int(dk)(d\varepsilon)G_{R}\left(\varepsilon+\frac{\omega}{2},k\right)G_{A}\left(\varepsilon-\frac{\omega}{2},k\right),
\end{equation}
which yields Eq.~\eqref{eq:B-omega} of the main text. Further, the ``current'' $\hat{{\cal J}}_{t}$ in Eq.~\eqref{eq:app:St2} is defined as
\begin{equation}
\hat{{\cal J}}_{t}=\hat{{\cal J}}_{t}^{(+)}-\hat{{\cal J}}_{t}^{(-)},\quad\text{with}\quad\hat{{\cal J}}_{t}^{(\pm)}=i\hat{{\cal V}}_{\pm}^{\dagger}\partial_{t}\hat{{\cal V}}_{\pm}
\end{equation}

For maximal misalignment of the measurements, $\theta = \pi$, we have 
${\cal B}(\omega) \propto \omega^{-1/2}$, see 
Eq.~\eqref{eq:B-omega}, and thus 
${\cal B}(t) \propto t^{-1/2}$ for the kernel in Eq.~\eqref{eq:app:St2}. We have thus a NLSM effective theory with a long-range coupling decaying as a power law. 
This non-locality of the NLSM action implies the emergence of superdiffusive L{\'e}vy flights.
If one expands the matrix $\hat{U}$  with respect to its deviations from the saddle point $\hat{\mathbb{I}}$, i.e., $\hat{U} = \exp(i \hat{\Phi}) \approx \hat{\mathbb{I}} + i \hat{\Phi}$ with $\Phi \ll 1$, then the lowest-order (quadratic in $\hat{\Phi}$)  terms in Eq.~\eqref{eq:app:St2} read:
\begin{equation}
    S[\hat{\Phi}]=\frac{1}{4}\int dx\,\Tr\left\{D\int_{-\infty}^{0}dt\,\left[\partial_{x}\hat{\Phi}(x,t)\right]^{2}
    +\int_{-\infty}^{0}dt_{1}dt_{2}\,{\cal B}(t_{1}-t_{2})\,\partial_{t}\hat{\Phi}(x,t_{1})\partial_{t}\hat{\Phi}(x,t_{2})\right\},
    \label{eq:S-Gaussian}
\end{equation}
where $D$ is the spatial diffusion coefficient given by Eq. (\ref{eq:DiffusionCoefficient}) of the main text.

This action yields the following propagator in the ``bulk'' (i.e., sufficiently far from the boundary in the time domain):
\begin{equation}
\left\langle \Phi_{ab}(\omega,q)\Phi_{cd}(-\omega,-q)\right\rangle =\left(\delta_{ad}\delta_{bc}-\frac{1}{R}\delta_{ab}\delta_{cd}\right)\mathcal{D}_{\Phi}(\omega,q),
\end{equation}
where the replica structure originates from tracelessness of generator $\hat{\Phi}$, and, for the half-filling case $n_0 = 1/2$:
\begin{equation}
\mathcal{D}_{\Phi}(\omega,q)=\frac{2}{\omega^{2}\Re{\cal B}(\omega)+Dq^{2}}\underset{\omega,q\to0}{\approx}\begin{cases}
\displaystyle \frac{2}{\omega^{2}/\gamma|\cos\frac{\theta}{2}|+Dq^{2}}, & \theta\neq\pi \,;\\[0.5cm]
\displaystyle \frac{2}{|\omega|^{3/2}/2 \gamma^{1/2}+Dq^{2}}, & \theta=\pi \,.
\end{cases}
\end{equation}

The following subtle point should be mentioned here.
The non-local Eq.~\eqref{eq:app:St2} is, strictly speaking, not gauge-invariant, even though it was obtained from the expansion of the manifestly gauge-invariant trace-log term. Indeed, a gauge transformation $\hat{{\cal V}}_{\pm}\mapsto\hat{{\cal V}}_{\pm}\hat{{\cal V}}$, which leaves matrix $\hat{U}$ unchanged, affects the current as $\hat{{\cal J}}_{t}\mapsto i\hat{{\cal V}}^{\dagger}\hat{{\cal J}}_{t}\hat{{\cal V}}$. 
In the conventional ``diffusive'' case ($\theta $ far from $\pi$ in our model), when ${\cal B}(t)$ can be replaced by the delta-function [i.e., $\mathcal{B}(\omega \to 0)$ is finite], the second order in the expansion of the trace-log acquires the following, gauge invariant, form:
\begin{equation}
S_{t}^{(2)}[\hat{Q}]\approx n_{0}(1-n_{0}){\cal B}(\omega=0)\int dxdt\Tr\left[\partial_{t}\hat{U}^{\dagger}\partial_{t}\hat{U}\right],
\end{equation}
characteristic of a diffusive NLSM.
In this case, higher-order terms of the expansion contain additional smallness in $\omega / \gamma \ll 1$. On the other hand,  in the case of the super-diffusion, with $\mathcal{B}(\omega \to 0)$ having a power-law divergence, a careful analysis reveals that higher-order terms do not contain such smallness and whole series has to be resummed in order to restore strict gauge invariance. These higher-order corrections to Eq.~\eqref{eq:app:St2}
have the structure
\begin{equation}
\label{eq:app:Stk}
S_{t}^{(k)}[\hat{Q}]\sim\sum_{s_{i} =\pm}\int dxdt_{1}\ldots dt_{k}{\cal B}_{s_{1}\dots s_{k}}^{(k)}(t_{1}-t_{2},\ldots,t_{1}-t_{k})\Tr\left[\hat{{\cal J}}_{t}^{(s_{1})}(x,t_{1})\ldots\hat{{\cal J}}_{t}^{(s_{k})}(x,t_{k})\right],\qquad k=3,4,\ldots\,,
\end{equation}
where the kernels ${\cal B}_{s_{1}\dots s_{k}}^{(k)}$ are homogeneous functions of fractional degree $-1/2$. When one expands in $\hat{\Phi}$ to the order $k$, only first $k$ terms in the series contribute, and their sum is gauge-invariant. In particular, to the Gaussian order, only the $k=2$ term \eqref{eq:app:St2} contributes, giving 
Eq.~\eqref{eq:S-Gaussian} of the main text, as explained above.

For weak monitoring (small $\gamma$), the Gaussian approximation, Eq.~\eqref{eq:S-Gaussian} of the main text, fully determines the density correlation function of the NLSM theory, with quantum corrections originating from higher terms being negligibly small. At the same time, one can ask whether quantum effects may lead to strong localization in the limit of large $\gamma$. If this would be the case, we would have a localization transition at some intermediate $\gamma$. A rigorous analytic answer to this question requires a careful renormalization-group analysis of our NLSM theory. We leave this for the future, providing here the following arguments. Renormalization-group analysis
of NLSM theories with power-law couplings was performed in Ref.~\cite{Mirlin1996transition} for a matrix NLSM, see also the early paper on a vector NLSM \cite{Brezin1976critical}. It was found in these works that, in the replica limit (in which the number of degrees of freedom is zero) and for sufficiently slowly decaying couplings, there is no infrared quantum corrections. As a consequence, there is no transition: the behavior of the correlation functions is the same as in the Gaussian (quasiclassical) approximation.  In fact, our NLSM is different in several aspects from those studied in Refs.~\cite{Mirlin1996transition,Brezin1976critical}: (i) it is of a different symmetry class, (ii) its action includes a series of terms  \eqref{eq:app:Stk}, (iii) the action is superdiffusive with respect to $t$ direction but conventional diffusive in $x$ direction. We expect, however, that these differences do not affect the conclusion about the absence of transition. An additional argument supporting this is the result of Ref.~\cite{Gattenloehner16Levy} where a similar theory (diffusion in one direction and superdiffusive in another direction) was obtained for a problem of transport in a 2D disordered system with a special type of disorder. It was found in Ref.~\cite{Gattenloehner16Levy}  from the inspection of a weak-localization correction (and supported by numerics there) that there are only finite quantum corrections but no localization transition. 

We thus argue that the superdiffusive behavior (at $\theta=\pi$) obtained from the Gaussian approximation to our NLSM holds also for large $\gamma$, and there is no localization transition in this model. This is supported also by our numerical results shown in the main text of the paper. 

\subsection{Boundary conditions}

Finally, we relate the derived NLSM field theory to observable quantities \cite{Poboiko2025}. The theory is defined on the semi-axis $t < 0$ in the time domain, whereas at $t = 0$ one has to introduce boundary conditions. We define the partition function with arbitrary boundary conditions as follows:
\begin{equation}
Z_{A}[\hat{U}_{0}]=\int{\cal D}\hat{U}\,\exp\left(-S[\hat{U}]\right),\quad\text{subject to}\quad\begin{cases}
\hat{U}(x\in A,t=0)=\hat{U}_{0}\oplus\hat{\mathbb{I}}_{R-N}\\
\hat{U}(x\notin A,t=0)=\hat{\mathbb{I}}_{R}
\end{cases},
\label{eq:app:Z}
\end{equation}
where the matrix $\hat{U}_0$ has a size $N \times N$ with $N \leq R$.

Both the entanglement entropy and the charge generating function for arbitrary region $A$ can be then expressed via such partition function as follows:
\begin{align}
S_{A}&\equiv-\overline{\Tr\left(\hat{\rho}_{A}\ln\hat{\rho}_{A}\right)}=-\lim_{N\to1}\frac{\ln Z[\hat{\mathbb{T}}_{N}]}{N-1},\\
\chi_A(\lambda)&\equiv\overline{\ln\left\langle e^{i\lambda\left(\hat{N}_{A}-\left\langle \hat{N}_{A}\right\rangle \right)}\right\rangle }=\lim_{N\to1}\frac{\ln Z[e^{i\hat{\lambda}_{N}}]}{N-1},
\label{eq:app:chi}
\end{align}
which involve auxiliary matrices of the following structure (shown here for $N = 3$):
\begin{equation}
\label{eq:app:BC}
\hat{\mathbb{T}}_{N}=\begin{pmatrix}0 & -1 & 0\\
0 & 0 & -1\\
1 & 0 & 0
\end{pmatrix},\quad e^{i\hat{\lambda}}=\begin{pmatrix}e^{i\lambda} & 0 & 0\\
0 & e^{i\lambda} & 0\\
0 & 0 & e^{-i(N-1)\lambda}
\end{pmatrix}.
\end{equation}

Formally, such a procedure determines the fluctuation of charge of fermions $d$, as well as entanglement entropy calculated in the basis of fermions $d$ rather than original fermions $c$. However, since the relation between $c$-fermions and $d$-fermions is local (involves two adjacent sites only) and we are interested in the behavior for a large size of the region $A$, this difference is not essential since it may only give corrections proportional to the surface area of the region $A$, i.e. $\sim O(\partial A)$, i.e., a constant of order unity for 1D systems studied here.

The entanglement entropy can be alternatively expressed through the statistics of fluctuations of charge via the Klich-Levitov relation:
\begin{equation}
\chi_{A}(\lambda)\equiv\sum_{n=0}^{\infty}\frac{(i\lambda)^{n}}{n!}{\cal C}_{A}^{(n)}\quad\Longrightarrow\quad S_{A}=\sum_{n=1}^{\infty}2\zeta(2n){\cal C}_{A}^{(2n)},
\end{equation}
where ${\cal C}_A^{(n)}$ is the $n$-th cumulant of charge. 

\begin{figure}
    \centering
    \includegraphics[width=0.5\columnwidth]{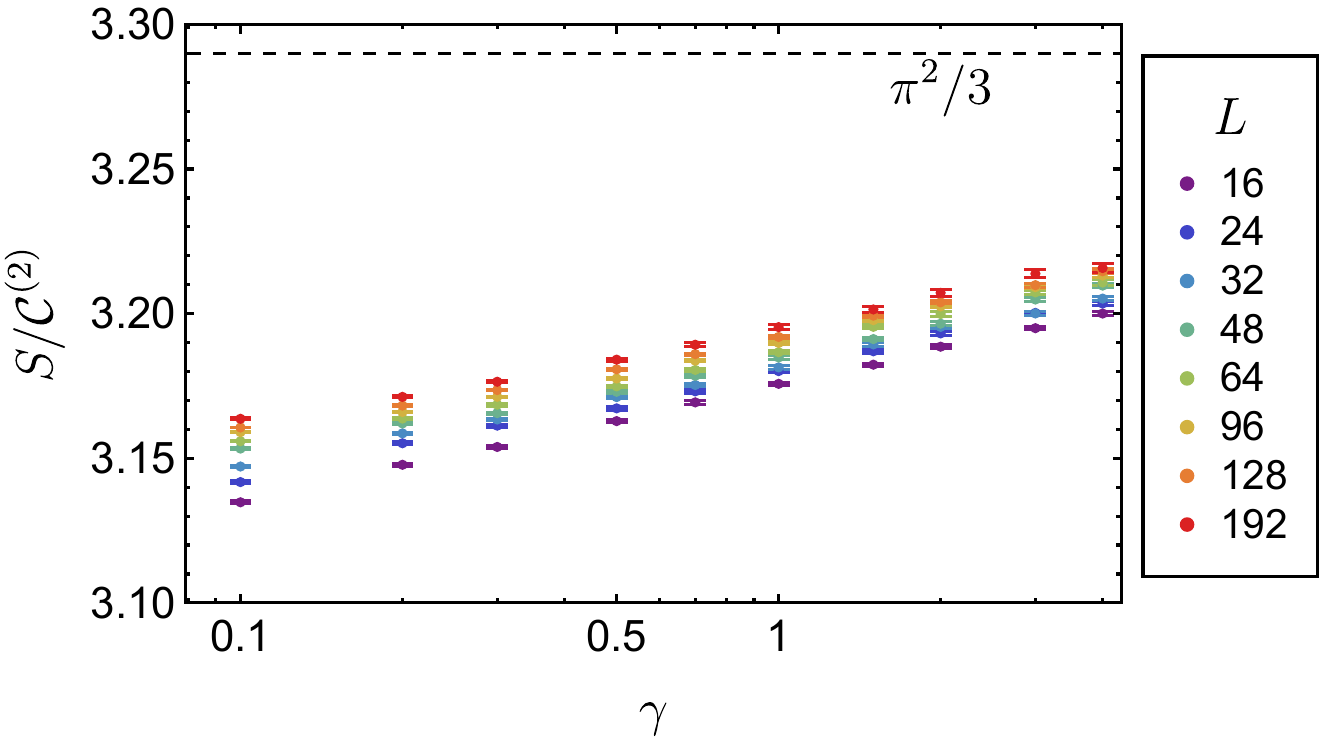}
    \caption{Numerical results for the ratio $S/\mathcal{C}^{(2)}$ for a subsystem of length $L/2$ at $\theta = \pi$, for various monitoring strength $\lambda$ and system size $L$. The dashed line shows the value of $\pi^2/3$.}
    \label{fig:S_over_C2}
\end{figure}

The quadratic expansion \eqref{eq:S-Gaussian} allows us to calculate only the second cumulant ${\cal C}_A^{(2)}$, or, equivalently, the pair density correlation function $C(x-y)$ in the Gaussian approximation controlled by a large value of the ``dimensionless conductance'' (inverse of which controls the magnitude of quantum corrections). However, we have seen earlier \cite{Poboiko2023a,Poboiko2023b} for a model with conventional density monitoring, the second cumulant is sufficient to determine the entanglement entropy with very good precision in all regimes.
We have checked numerically that this holds also for the present model. In Fig.~\ref{fig:S_over_C2}, the ratio of the entanglement entropy to the second cumulant, $S/\mathcal{C}^{(2)}$, in our superdiffusive model ($\theta=\pi$) is shown for various system sizes and various monitoring strength. It is seen that this ratio is very close (within a few percent) to $2\zeta(2) = \pi^2/3$, 
i.e., Eq.~\eqref{eq:ratio-S-C2} of the main text holds with excellent accuracy. 

\section{Saddle point approximation: \ Density correlation function and second charge cumulant}
\subsection{Action}

In this Section of the Supplemental Material, we calculate the density-density correlation function $C(q)$ and charge fluctuations ${\cal C}_A^{(2)}$ in the saddle-point approximation. This requires finding the minima of action \eqref{eq:S-Gaussian} subject to the following boundary conditions:
\begin{equation}
\hat{\Phi}(x,t=0)=\begin{pmatrix}\lambda(x) & 0 & 0\\
0 & \lambda(x) & 0\\
0 & 0 & -(N-1)\lambda(x)
\end{pmatrix},
\end{equation}
which follow from Eqs.~(\ref{eq:app:Z},\ref{eq:app:chi},\ref{eq:app:BC}), and where we have kept the density source $\lambda(x)$ arbitrary. The charge fluctuations in the given region $A$ correspond to the specific choice $\lambda(x\in A)=\lambda$ and zero otherwise. Given the diagonal form of boundary conditions, the solution for the saddle-point equations can also be sought in the diagonal form:
\begin{equation}
\hat{\Phi}(x,t)=\varphi(x,t)\cdot\begin{pmatrix}1 & 0 & 0\\
0 & 1 & 0\\
0 & 0 & -(N-1)
\end{pmatrix},
\end{equation}
with the scalar function $\varphi(x,t)$. The action \eqref{eq:S-Gaussian} with this Ansatz reduces to
\begin{equation}
S[\hat{\Phi}]=N(N-1)S_{0}[\varphi].
\end{equation}
The generating function \eqref{eq:app:chi} in the saddle-point approximation then yields $\chi[\lambda]=-S_{0}[\varphi]$, with the action calculated on the saddle-point solution.

With the suitable choice of units of time, the action $S_0$ can be brought to the following form:
\begin{equation}
\label{eq:app:S-scalar}
S_0[\varphi]=\frac{g_{0}}{2}\int dx\left[\int_{-\infty}^{0}dt\,\big(\partial_{x}\varphi(x,t)\big)^{2}+\int_{-\infty}^{0}dt_{1}\int_{-\infty}^{0}dt_{2}\,L(t_{1}-t_{2})\,\partial_{t}\varphi(x,t_{1})\partial_{t}\varphi(x,t_{2})\right],\quad L(\omega)=|\omega|^{\alpha-2},
\end{equation}
subject to the boundary conditions $\varphi(x,t=0)=\lambda(x)$. Here, $\alpha = 2$ corresponds to standard diffusion and $\alpha = 3/2$ corresponds to the model under consideration. The constant $g_0$ will be specified below. In the present calculation, we will consider arbitrary $\alpha \in (1, 2]$. 

\subsection{Wiener-Hopf problem}
We perform a spatial Fourier transformation and seek for the solution in the form:
\begin{equation}
\varphi_{q}(t)=\lambda_{q}\psi\left(\tau\equiv q^{2/\alpha}t\right),
\end{equation}
with a single dimensionless function $\psi(\tau)$; the saddle-point equation is then equivalent to
\begin{equation}
\label{eq:app:WienerHopfEquation}
\psi(\tau)-\frac{\partial^2}{\partial \tau^2}\int_{-\infty}^{0}d\tau^{\prime}L(\tau-\tau^{\prime})\psi(\tau^{\prime})=\frac{\partial L(\tau)}{\partial\tau},\quad \psi(0) = 1.
\end{equation}
The action calculated on such a saddle-point configuration reads:
\begin{equation}
\label{eq:app:S-C}
S_0[\varphi]=\frac{1}{2}\int(dq)C(q)\lambda_{q}\lambda_{-q},\quad C(q)=c_{0} g_{0}|q|^{2\left(1-1/\alpha\right)},\quad c_{0}=\int_{-\infty}^{0}d\tau\psi(\tau).
\end{equation}
This form of the saddle point action then implies that the function $C(q)$ then identically yields the correlation function of densities.

To solve Eq.~\eqref{eq:app:WienerHopfEquation}, we employ the Wiener-Hopf method. We continue this equation to the entire line $\tau \in \mathbb{R}$ and seek the solution in the form $$\psi(\tau)=\psi_{+}(\tau)+\psi_{-}(\tau),$$ where $\psi_{+}(\tau<0)=\psi_{-}(\tau>0)=0$ [that is, the function $\psi_{+}(\tau)$ is retarded and the function $\psi_{-}(\tau)$ is advanced]. 
Equation \eqref{eq:app:WienerHopfEquation} can then be solved using the Fourier transform:
\begin{equation}
\label{eq:app:WienerHopf:Fourier}
\psi_{+}(\omega)+\left(1+\omega^{2}L(\omega)\right)\psi_{-}(\omega)=-i\omega L(\omega).
\end{equation}

As the next step, we perform the Wiener-Hopf factorization:
\begin{equation}
\frac{1}{1+\omega^{2}L(\omega)}=K(\omega)K^{\ast}(\omega),
\end{equation}
with function $K(\omega)$ being analytic in the upper complex half-plane (and, thus, $K^\ast(\omega)$ is analytic in the lower complex half-plane). We obtain:
\begin{equation}
\ln K(\omega)=\int_{-\infty}^{+\infty}\frac{dz}{2\pi i}\frac{\ln(1+z^{2}L(z))}{\omega+i0-z}=\omega\int_{0}^{+\infty}\frac{dz}{\pi i}\frac{\ln(1+z^{\alpha})}{(\omega+i0)^{2}-z^{2}}.
\end{equation}
Additionally, we perform the subsequent factorization:
\begin{equation}
-i\omega K(\omega)L(\omega)\equiv f_{+}(\omega)+f_{-}(\omega),
\end{equation}
with retarded (advanced) functions $f_{\pm}(\omega)$, with the help of the following integral representations:
\begin{equation}
f_{\pm}(\omega)=\mp\int_{-\infty}^{\infty}\frac{dz}{2\pi}\frac{zL(z)K(z)}{z-\omega\mp i0}.
\end{equation}
Performing both factorizations, we reduce Eq.~\eqref{eq:app:WienerHopf:Fourier} to
\begin{equation}
K(\omega)\psi_{+}(\omega)+\frac{\psi_{-}(\omega)}{K^{\ast}(\omega)}=f_{+}(\omega)+f_{-}(\omega)\ \Rightarrow \ \begin{cases}
\psi_{-}(\omega)=f_{-}(\omega)K^{\ast}(\omega)\\
\psi_{+}(\omega)=f_{+}(\omega)/K(\omega)
\end{cases}.
\end{equation}
Finally, we note that the number $c_{0}$ is given by
\begin{equation}
\label{eq:app:C}
c_{0}=\int_{-\infty}^{0}d\tau\,\psi(\tau)=\psi_{-}(\omega=0) = f_{-}(0) = \int_{0}^{\infty}\frac{d\omega}{\pi}L(\omega)\Re K(\omega).
\end{equation}

One can perform a transformation of the integration contour, such that the integration in Eq.~\eqref{eq:app:C} runs along the branch cut [required to define $L(\omega)$] parallel to the imaginary axis, arriving at another representation for $c_{0}$:
\begin{equation}
c_{0}=\frac{1}{\pi}\sin\frac{\pi\alpha}{2}\int_{0}^{\infty}\frac{dx}{x^{2-\alpha}}g(x),\qquad\ln g(x)=-\int_{0}^{+\infty}\frac{dy}{\pi}\frac{\ln(1+x^{\alpha}y^{\alpha})}{1+y^{2}}.
\label{C-WH-int}
\end{equation}
Remarkably, the integration in Eq.~(\ref{C-WH-int}) can be performed analytically. Identical integrals appear in the analysis of the extrema of L{\'e}vy-stable processes. Such integrals were studied extensively in Ref.~\cite{Kuznetsov2011}, where the Mellin transform of $g(x)$ was expressed in terms of the Barnes $G$-functions:
\begin{equation}
\Phi(s,\alpha)\equiv\int_{0}^{\infty}dx\,x^{s-1}g(x)=\frac{\alpha^{-s}}{\sqrt{\pi}}\Gamma\left(s\right)\Gamma\left(1-\frac{s}{\alpha}\right)\frac{G_{\alpha}\left({\alpha}/{2}+1+s\right)}{G_{\alpha}\left({\alpha}/{2}-s\right)}\frac{G_{\alpha}\left(\alpha-s\right)}{G_{\alpha}\left(\alpha+s\right)}.
\end{equation}
Utilizing the recurrence relations
\begin{align}
G_{\alpha}(z+1)&=\Gamma\left(\frac{z}{\alpha}\right)G_{\alpha}(z),
\\
G_{\alpha}(z+\alpha)&=(2\pi)^{(\alpha-1)/2}\alpha^{-z+1/2}\,\Gamma(z)G_{\alpha}(z),
\end{align}
we obtain:
\begin{equation}
\Phi(\alpha-1,\alpha)=\frac{\pi}{\sin\frac{\pi\alpha}{2}\sin\frac{\pi}{\alpha}}  \quad \Longrightarrow \quad c_{0}=\frac{\sin\frac{\pi\alpha}{2}}{\pi}\Phi(\alpha-1,\alpha)=\frac{1}{\sin\frac{\pi}{\alpha}}.
\label{eq:app:C-res}
\end{equation}

Finally, we provide values of $g_0$, which enter \eqref{eq:app:S-scalar}, and which correspond to our problem:
\begin{align}
\text{diffusion:}&&\alpha&=2&g_{0}&=\ell_{0}/2\sqrt{\left|\cos\frac{\theta}{2}\right|},\\
\text{superdiffusion:}&&\alpha&=3/2&g_{0}&=\ell_{0}^{2/3}/2^{5/3}.
\label{eq:app:g0}
\end{align}
Combining this with Eqs.~(\ref{eq:app:S-C},\ref{eq:app:C-res}) yields Eq.~\eqref{eq:C-q} of the main text.

\subsection{Second cumulant}
We finally discuss the behavior of the second cumulant of charge, which follows from the obtained form of the density correlation function $C(q)$, for the case when the size $\ell$ of the subsystem might be of the order of the size of the whole system, $r \equiv \ell / L \in [0, 1]$. For such a case, one has to take into account momentum quantization $q_n = 2 \pi n / L$, which holds for the finite system. Then, Eq.~\eqref{eq:C2} yields:
\begin{equation}
\label{eq:app:CA-Cq}
{\cal C}_{A}^{(2)}=\frac{1}{L}\sum_{q}C(q)\left(\frac{2}{q}\sin\frac{q\ell}{2}\right)^{2} \approx g_{0}\ell^{2/\alpha-1}c_{\alpha}(r),
\end{equation}
with dimensionless function $c_{\alpha}(r) = O(1)$:
\begin{equation}
c_{\alpha}(r)=8c_{0}r\sum_{n=1}^{\infty}\frac{\sin^{2}(\pi nr)}{(2\pi nr)^{2/\alpha}}=\frac{4r}{(2\pi r)^{2/\alpha}\sin(\pi / \alpha)}\left[\zeta\left(\frac{2}{\alpha}\right)-\Re\text{Li}_{2/\alpha}\left(e^{2i\pi r}\right)\right],
\end{equation}
with the polylogarithm function $\text{Li}_{2/\alpha}(z)$. In the two important cases---the infinite system limit $r \to 0$ and the half-system bipartition case $r = 1/2$---it yields:
\begin{equation}
c_{\alpha}(r)=\begin{cases}
\displaystyle{\frac{4}{\pi\alpha}\Gamma\left(-\frac{2}{\alpha}\right)}, & r\to0\\[0.5cm]
\displaystyle{\frac{4\left(4^{1/\alpha}-1\right)}{(2\pi)^{2/\alpha}\sin(\pi / \alpha)}\zeta\left(\frac{2}{\alpha}\right)}, & r=1/2.
\end{cases}
\end{equation}
For $\alpha = 3/2$ it reduces to:
\begin{equation}
c_{3/2}(r)=\begin{cases}
2.58617, & r\to0\\
2.18024, & r=1/2,
\end{cases}
\end{equation}
This, together with Eq.~\eqref{eq:app:g0}, leads to scaling given by Eq.~\eqref{eq:S-scaling} of the main text, and gives the analytical value of the numerical prefactor in the scaling of the second cumulant.

\section{Quantum corrections for \texorpdfstring{$\theta = \pi$}{theta = pi}}
\label{sec:QuantumCorrections}

In this section, we discuss the structure of quantum corrections for the NLSM with an anisotropic super-diffusive kernel, Eq.~\eqref{eq:S-Gaussian}. These corrections develop on top of the ``semiclassical'' behavior corresponding to the Gaussian approximation. Note that, in addition to the quantum corrections, the actual behavior for finite systems also involves ballistic corrections (see Section \ref{sec:app:ballistic} below), i.e., those corresponding to the short-scale ``semiclassical'' physics 
(coming from scales shorter or of the order of the mean free path $\ell_0$)
not captured by the NLSM. 
The analysis in the present discussion is particularly relevant to the measurement-only limit, $\gamma\to \infty$, where a crossover between AIII and BDI universality classes occurs. Since the semiclassical behavior (including ballistic corrections) is 
not sensitive to the symmetry class,
we ignore ballistic effects and focus entirely on the quantum corrections in this section.

Following the general approach for anisotropic systems developed in Ref.~\cite{Wolfle1984} (see also Ref.~\cite{Gattenloehner16Levy}), we expect that the renormalization of the diffusion kernel is also  anisotropic and takes the following form in terms of the integral over $\boldsymbol{q}=(q,\omega)$, describing the return probability in the $(x,t)$ plane:
\begin{equation}
\delta D_{\alpha\alpha}\propto-\int (d^2 \boldsymbol{q})\frac{D_{\alpha\alpha}(\boldsymbol{q})}{\sum_{\beta}D_{\beta\beta}(\boldsymbol{q})q_{\beta}^{2}}.
\end{equation}
Here, the diffusion coefficient in the spatial direction $D_{xx} \equiv D = \const$ and frequency-dependent diffusion coefficient in the temporal direction:
\begin{equation}
D_{tt}(\omega) \equiv \mathcal{B}(\omega) \equiv b / \sqrt{|\omega|},
\label{eq:app:Dtt}
\end{equation}
see Eq.~\eqref{eq:B-omega} of the main text.

This form of the correction implies that the correction to the spatial diffusion coefficient is infrared-finite, being dominated by the contribution of ultraviolet scales  (i.e., it is non-universal), while the scales of order of the system size $L\gg \ell_0$ give a non-divergent subleading contribution:
\begin{equation}
\delta D_{xx}\sim-D\int\frac{d\omega dq}{Dq^{2}+b|\omega|^{3/2}}\sim-\left(\frac{D}{b}\right)^{2/3}\int_{\sim L^{-1}}^{\sim\ell_{0}^{-1}}dq |q|^{-2/3}\sim-\left(\frac{D}{b}\right)^{2/3}\left(\ell_{0}^{-1/3}-L^{-1/3}\right).
\label{eq:DCorrection}
\end{equation}
Such a structure of the correction is familiar from the Anderson localization theory, where the weak-localization correction in conventional (isotropic, normal diffusion) 3D situation is also governed by the ultraviolet scale and thus non-singular at $L\to \infty$.
The relative magnitude of the correction is 
\begin{equation}\frac{|\delta D_{xx}|}{D_{xx}} \sim
\begin{cases}
\gamma / J, &\quad  \gamma / J \ll 1,\\
1, & \quad \gamma / J \gtrsim 1.
\end{cases}
\label{eq:S65}
\end{equation} 
At the same time, the correction to the temporal coefficient is less singular at $\omega\to 0$ than its bare value \eqref{eq:app:Dtt}:
\begin{equation}
\delta D_{tt}\sim-\int\frac{b|\omega|^{-1/2}d\omega dq}{Dq^{2}+b|\omega|^{3/2}}\sim-\left(\frac{b}{D}\right)^{1/2}\int d\omega|\omega|^{-5/4}\sim-\left(\frac{b}{D}\right)^{1/2}|\omega|^{-1/4}.
\end{equation}
According to the superdiffusive scaling $|\omega|\sim(D/b)^{2/3}|q|^{4/3}$ and $|q| \sim L^{-1}$, this leads to the following relative correction:
\begin{equation}
\frac{|\delta D_{tt}|}{D_{tt}}\sim\left(\frac{1}{b^{2}D}\right)^{1/3}L^{-1/3}
\label{eq:app:Dtt:correction}
\end{equation}
This implies that, in the leading order, the correction to $D_{xx}$ is more important; however, the $L$-dependent contribution to $D_{tt}$ has the same scaling as the subleading $L$-dependent contribution to $D_{xx}$.

Taking into account the renormalization of $D_{xx}$ in the density correlation function,
\begin{equation}
C(q)\sim\int\frac{D_{xx}(\boldsymbol{q})D_{tt}(\boldsymbol{q})q^{2}}{D_{xx}(\boldsymbol{q})q^{2}+D_{tt}(\boldsymbol{q})\omega^{2}}d\omega\sim\left(Db^{2}q^{2}\right)^{1/3},
\end{equation}
one observes that the relative corrections to $D$ imply similar relative corrections to the second cumulant $\mathcal{C}_{A}^{(2)}$ and the entanglement entropy $S_A$. Importantly, the fractal scaling 
$S\sim s L ^{1/3}$ persists in the limit $L\to \infty$ when quantum corrections are taken into account: it is only the prefactor $s$ in this scaling that slowly crosses over from its Gaussian value at $L\sim \ell_\ast$ to the value at $L\to \infty$, which includes finite quantum corrections. The exact value of the crossover scale $\ell_\ast$ is expected to be parametrically the same as the mean free path $\ell_0$. However, in our simulations, we observe that the crossover is very slow and happens at a scale that is numerically larger ($\ell^\ast / \ell_0 \gtrsim 30$). For this reason, we introduce additional notation for the crossover scale.

\begin{figure}
    \centering
    \includegraphics[width=0.6\linewidth]{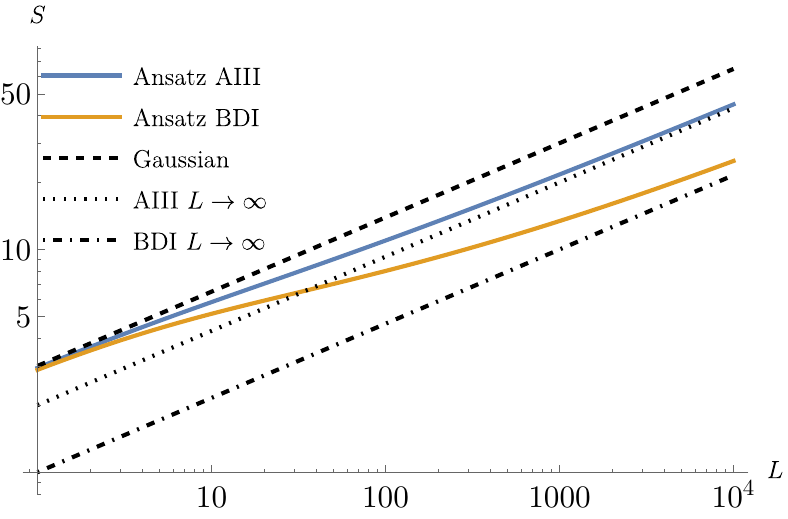}
    \caption{Illustration of the crossover behavior of $S(L)$ given by Eq.~\eqref{eq:SLSpeculation} with the following 
    choice of parameters: $s_0 = 3$,\ $\delta s^{\text{(AIII)}} = 1$,\ $\delta s^{\text{(BDI)}} = 2$. For the sake of comparison, it is assumed that both classes are realized at the same finite value of $\gamma\sim 1$, such that the Gaussian results for these classes coincide (dashed line), while the quantum corrections differ by the factor of 2, yielding different prefactors at $L\to \infty$ (dotted and dash-dotted lines for classes AIII and BDI, respectively). Strictly speaking, in our model, class BDI is only realized at $\gamma=\infty$, for which case $s_0$ is reduced and the role of quantum corrections is further enhanced. 
    Note that the crossover spans several orders of magnitude, and the apparent power-law exponent appears smaller in the intermediate regime. At the same time, for not too large values of $\gamma$, the crossover for the AIII class takes place between two rather close values of the prefactor, as the blue curve in the plot shows. (For small $\gamma$, the relative effect of the quantum correction is still smaller.) }
    \label{fig:SCSpeculation}
\end{figure}

Let us now discuss how quantum corrections manifest themselves in finite-size numerics. 
In view of Eqs.~\eqref{eq:DCorrection}, \eqref{eq:app:Dtt:correction}, we 
propose the following Ansatz for the crossover between the small-$L$ and large-$L$ asymptotics of the entropy:
\begin{equation}
S(L)=\left\{s_{0} - \delta s\, \left[1-\left(\frac{\ell_\ast}{L+\ell_\ast}\right)^{1/3}\right]\right\}L^{1/3},
\label{eq:SLSpeculation}
\end{equation}
which is illustrated in Fig.~\ref{fig:SCSpeculation}. 
At small scales $L \lesssim \ell_\ast$, this Ansatz yields the ``Gaussian'' result $s_0 L^{1/3}$, whereas at larger $L$, it incorporates quantum corrections $\propto \delta s$ with the slow power-law dependence on $L$ consistent with Eqs.~\eqref{eq:DCorrection}, \eqref{eq:app:Dtt:correction}. The prefactor $s$ in $L\to \infty$ limit is given by $s_0-\delta s$. 

One can observe in Fig.~\ref{fig:SCSpeculation} that, because of the fractal $L$-scaling of both the Gaussian result and the quantum correction, with a relatively small power $1/3$, the approach of $S(L)$ to its true thermodynamic asymptotic behavior is very slow and can span several orders of magnitude (even though the equation itself does not contain any large parameters). Furthermore, at intermediate scales that are accessible to our numerical simulations, such behavior would appear like a power-law scaling with a slightly reduced (compared to 1/3) exponent.

The ``bare'' value of the prefactor, $s_0$, is given by the Gaussian approximation discussed above, and is large for small $\gamma$: $s_0 \propto (J/\gamma)^{2/3}$. The quantum correction $\delta s$ is negative (at least, at  one-loop order) and is dominated by the short-distance physics. According to Eq.~\eqref{eq:S65}, the relative effect of the quantum correction is more prominent for larger values of $\gamma$. 

Furthermore, for the special point $\gamma = \infty$, an additional effect comes into play, because the system belongs to a different symmetry class at this point: BDI instead of AIII. The one-loop weak-localization correction in the class BDI is exactly twice larger compared to AIII. This is illustrated in Fig.~\ref{fig:SCSpeculation} by the comparison of the results for the two symmetry classes (blue and orange curves) at the same values of the bare parameters with $\gamma\sim 1$, such that the Gaussian asymptotics are the same for the two curves. One sees that the effect of quantum corrections appears to be significantly stronger for the BDI curve in this plot. In the measurement-only limit (where BDI symmetry is actually realized in our problem), the prefactor in the Gaussian asymptotics is of order unity, which makes the relative effect of quantum corrections for $\gamma = \infty$ even more prominent. 

We thus argue that the apparent deviations from the $L^{1/3}$ scaling observed in our numerical simulations in the case of $\gamma = \infty$ (see inset in Fig.~\ref{fig:ent_collapse} of the main text) is due to the intermediate-scale effect of quantum corrections in the symmetry class BDI, taken in the strong-coupling limit of the theory. At the same time, the influence of quantum corrections on the entanglement entropy for other curves in Fig.~\ref{fig:ent_collapse} of the main text (class AIII with not-too-large values of $\gamma$) is relatively weak, so that it is hard to distinguish between the Gaussian and thermodynamic-limit $L^{1/3}$ asymptotics for those curves.  

\section{Numerical details}
\label{sec:NumericalDetails}

Numerical simulation is done using the efficient representation of Gaussian states for free fermionic systems with U(1) symmetry introduced in Ref.~\cite{Cao2019a}. In short, a state of $N$ fermions on $L$ sites is described using an $L\times N$ complex matrix $U$, where each column represents a single-particle mode,
\begin{equation}
    |\psi\rangle = \prod_{n = 1}^N \left(\sum_{i = 1}^L U_{i n} c^\dagger_i\right) |0\rangle,
\end{equation}
where $c^\dagger_i$ is the fermionic creation operator on site $i$ and $|0\rangle$ is the vacuum state.
The correlation matrix can be retrieved as $\mathcal{G}^\ast = U U^\dagger$. We initialize the system by randomly placing $N$ particles on $L$ sites.

Evolution through the stochastic Schr\"odinger equation in Eq.~\eqref{eq:SSE} is equivalent to the following change in $U$,
\begin{equation}
\label{eq:numerics:discrete-evolution}
  U' = e^{\mathbb{M}} e^{i\,dt\,\mathbb{H}} U,
\end{equation}
where $\mathbb{H}$ is the $L\times L$ single-particle Hamiltonian matrix, and $\mathbb{M}$ is the $L\times L$ representation of the measurement operator. The operator $U'$ is then normalized by taking its (thin) QR decomposition, and assigning the final $U'$ to be the matrix $Q$. For our model, 
\begin{equation}
    \mathbb{H}_{ij} = J ( \delta_{i,j+1} + \delta_{i,j-1}),
    \label{eq:numerics:ham}
\end{equation}
which corresponds to the hopping Hamiltonian. The measurement procedure consists of $L$ individual measurements of observables $M_k$, given by Eq.~\eqref{eq:measurement-operator} of the main text, which are defined on bonds $k$ connecting sites $k$ and $k+1$. The operators $M_k$ on adjacent bonds do not commute. The matrix $\mathbb{M}^{(k)}$ corresponding to the measurement operator $M_k$ reads:
\begin{equation}
\label{eq:numerics:krauss}
    \mathbb{M}_{i j}^{(k)} = A_k \delta_{k, i} \left( \delta_{i,j} \cos^2 \frac{\theta}{4} + \delta_{i+1,j+1} \sin^2 \frac{\theta}{4} + (\delta_{i,j+1} + \delta_{i+1,j}) \sin \frac{\theta}{4} \cos \frac{\theta}{4} \right),
\end{equation}
where
\begin{equation}
\label{eq:numerics:norm}
    A_k = (2 \langle M_k \rangle - 1) \gamma\, dt + d\xi_k^t
\end{equation}
is the coefficient representing the stochastic character of measurements, and
\begin{equation}
    \langle M_k \rangle = \sum_l \left| U_{k,l} \cos \frac{\theta}{4} + U_{k+1,l} \sin \frac{\theta}{4}\right|^2
\end{equation}
is the expectation value of the measurement operator $M_k$.

\begin{figure}
    \centering
    \includegraphics[width=0.9\columnwidth]{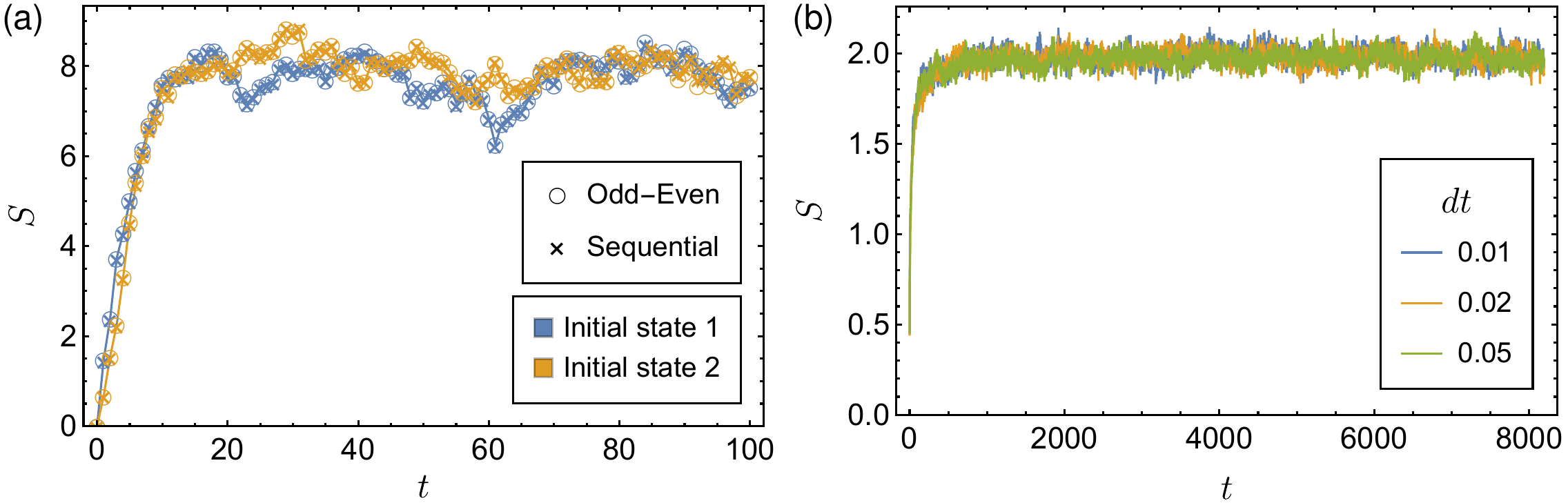}
    \caption{(a)~Early-time dependence of half-chain entanglement entropy $S$ of a single trajectory for different random initial states and different sequences of measurements in the implementation of the continuous monitoring, with $L=64,\,dt=0.02,\,\gamma=0.2,\,\theta=\pi$. (b)~Early-time dependence of the trajectory-averaged half-chain entanglement entropy $S$ for different values of $dt$, showing agreement in the steady-state value, with $L=64,\,J=0,\,\gamma=1$ (measurement-only model), $\theta=\pi$.}
    \label{fig:trajectories}
\end{figure}

We trotterize the measurement step, where we first calculate the action of the measurements on even bonds, followed by normalization, and then the measurements on odd bonds, again followed by normalization, leading to the following stochastic evolution
\begin{equation}
    U(t+dt) = \mathcal{N}\!\left[\exp\Big(\sum_{k\text{ odd}} \mathbb{M}^{(k)}\Big)\, \mathcal{N}\!\left[\exp\Big(\sum_{k\text{ even}}\mathbb{M}^{(k)}\Big) e^{i dt \mathbb{H}} U(t)\right]\right],
    \label{eq:numerics:trotter}
\end{equation}
where $\mathcal{N}[\cdot]$ is the normalization procedure described above.
Note that the order of measurements within the small time step $dt$ becomes immaterial in the limit $dt \to 0$, and here we have chosen a sequence that minimizes the number of normalizations (QR decompositions) per time step. Comparison between this odd-even sequence and a protocol where the measurements are performed sequentially from left to right within the time step $dt$ is shown in Fig.~\ref{fig:trajectories}(a), for two different initial states. We can see that odd-even and sequential protocols yield virtually indistinguishable results for a given initial state, which means that the order within the time step $dt \to 0$ is indeed immaterial. We also see that the curves for different initial states fluctuate around the \emph{same steady-state value} at long times, which illustrates that our results are independent of the initial state. We also analyze the measurement-only limit of the model (designated as $\gamma = \infty$), which is defined by first taking the continuous-time limit $dt\to 0$, followed by the strong monitoring limit $\gamma/J\to\infty$. In practice, we set $J = 0$ and $\gamma = 1$, while ensuring that the time discretization $dt$ remains sufficiently small. We have checked that the steady-state properties are independent of discretization step for values of $dt$ used in the main text, as illustrated in Fig.~\ref{fig:trajectories}(b).

The particle-number covariance $G_{AB}$ can be calculated directly from the correlation matrix $\mathcal{G}$,
\begin{equation}
  G_{AB} = \sum_{\substack{i\in A\\j\in B}} |\mathcal{G}_{i j}|^2,
\end{equation}
as well as the pair density correlation function,
\begin{equation}
  C_{i j} = \mathcal{G}_{i j} \delta_{i j} - \mathcal{G}_{i j} \mathcal{G}_{j i}.
\end{equation}
The corresponding quantity in the momentum space, $C(q)$, is obtained by performing a fast Fourier transform on the rows of matrix $C_{i j}$.
The entanglement entropy $S$ of a region $A$ is calculated by diagonalizing the sub-matrix of $\mathcal{G}$, with indices corresponding to $A$, resulting in eigenvalues $\lambda_i$. Then,
\begin{equation}
  S = -\sum_{i}\Big(\lambda_i \ln \lambda_i + (1-\lambda_i) \ln (1-\lambda_i)\Big).
\end{equation}

\begin{figure}
    \centering
    \includegraphics[width=\columnwidth]{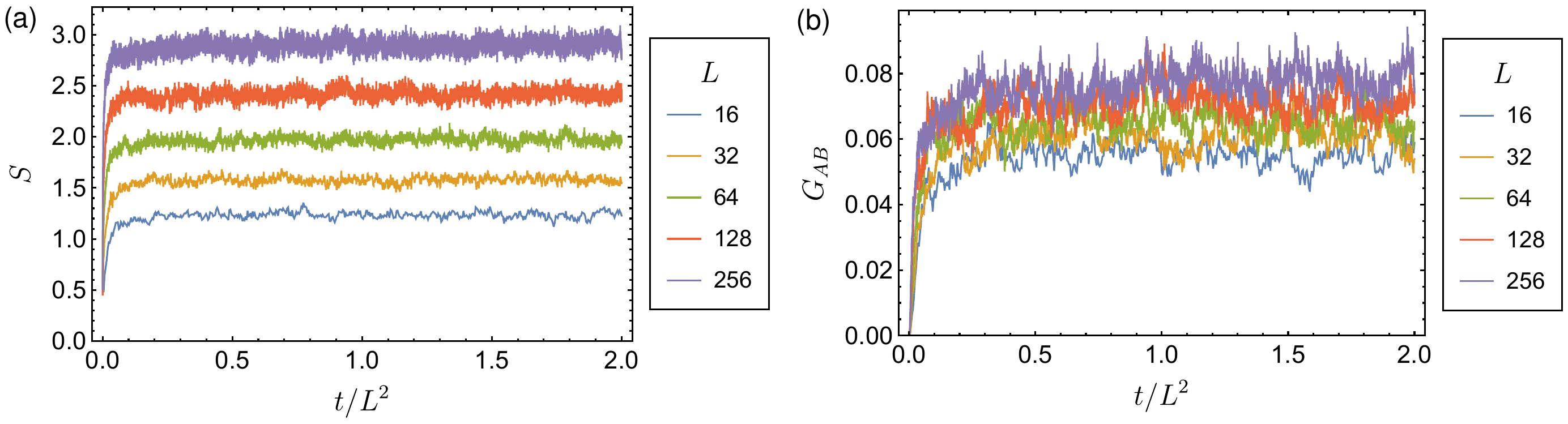}
    \caption{Early-time dependence of (a)~half-chain entanglement entropy $S$ and (b)~particle-number covariance $G_{AB}$ for the measurement-only model ($\gamma = \infty$) at $\theta=\pi$. The system is initialized at $t=0$ and then evolves due to monitoring.  }
    \label{fig:time_evol}
\end{figure}

Example results for the time dependence of the half-chain entropy $S$ and the particle-number covariance $G_{AB}$ are shown in Fig.~\ref{fig:time_evol}. We can see that numerically, the equilibration period, before the steady state is reached, is roughly $t_\text{equil} = L^2$. We average the results over the time range between $t_\text{equil}$ and $t_\text{equil} + 200$, and over 1000 random realizations (of the measurement and the initial state) for $L<320$ and 100 realizations for $L = 320$. We also find that the time discretization of $dt = 0.05$ accurately describes continuous evolution for $\gamma < 1$, while we use $dt = 0.02$ for $\gamma \ge 1$ and the measurement-only model.

\section{Numerical analysis of finite-size corrections}
\label{sec:app:ballistic}

In this Section, we perform numerical analysis of finite-size corrections to the scaling behavior of charge fluctuations $\mathcal{C}_A^{(2)}$ and the entanglement entropy $S(L)$ in the superdiffusive regime, $\theta= \pi$ and $L \gg \ell_0$.
Inspecting the numerical data for largest $\gamma$, such as $\gamma=2$, 3, and 4, and largest $L$ in Fig.~\ref{fig:ent_collapse}(b) of the main text, one sees that deviations from the asymptotic superdiffusive scaling $S(L) \propto L^{1/3}$ remain noticeable, even though the ratio $L / \ell_0$ reaches the values up to $\approx 400$.
We will first discuss the origin of these corrections, which clearly decay rather slowly with $L / \ell_0$, and then verify the analytical understanding by analyzing the numerical data.

The main source of the above corrections to the asymptotic superdiffusive scaling is in the short-scale (``ballistic'') behavior of correlations in the system. Such behavior (which is not captured by the long-wavelength NLSM theory) manifests itself in a modification of the density correlation function $C(x - x^\prime)$ at short scales $|x - x^\prime| \lesssim \ell_0$. Owing to particle number conservation (which implies nullification of $C(q=0) = 0$), the second cumulant of charge in an arbitrary region $A$ can be equivalently rewritten as
\begin{equation}
{\cal C}_{A}^{(2)}=\int_{x\in A}dx\int_{x^{\prime}\in B}dx^{\prime}C(x-x^{\prime}),
\end{equation}
where $B$ is the rest of the system. Modification of the short-scale behavior contributes to this integral only when both points $x$ and $x^\prime$ are near the boundary $\partial A$ of the region, so that finite-size corrections are expected to have an area-law form,   $|\delta{\cal C}_{A}^{(2)}|\sim \ell_0$. 
This physical argument can be confirmed by a more accurate analytical calculation. According to Eq.~\eqref{eq:C-q} of the main text, the Fourier transform $C(q)$ crosses over from the superdiffusive scaling $\propto |q \ell_0|^{2/3}$ to a ballistic scaling $\sim \const$ at $|q| \sim \ell_0^{-1}$. Thus, the superdiffusive approximation valid for $L \gg \ell_0$ 
\footnote{In the opposite limit $L \ll \ell_0$, the sum \eqref{eq:app:CA-Cq} is dominated by $q \sim L^{-1} \gg \ell_0^{-1}$. Substitution of the corresponding asymptotic behavior $C(q) \sim \const$ then yields a simple estimation $C_A^{(2)} \sim L$; the same estimation also holds for the entanglement entropy. This behavior corresponds to the dashed line in Fig.~\ref{fig:ent_collapse} of the main text.}, 
Eq.~\eqref{eq:app:CA-Cq}, overestimates the contribution from $q \gtrsim \ell_0^{-1}$, and such scales yield a (negative) correction, which can be estimated as (for $\ell = L/2$)
\begin{equation}
    |\delta{\cal C}_{A}^{(2)}|\sim \frac{1}{L}\sum_{|q|\gtrsim\ell_{0}^{-1}}|q\ell_{0}|^{2/3}\left(\frac{2}{q}\sin\frac{qL}{4}\right)^{2}\sim \ell_0.
\end{equation}
These arguments equally apply to charge fluctuations $C_{A}^{(2)}$ and to the half-system entanglement entropy $S(L)$, as these are directly related via Eq.~\eqref{eq:ratio-S-C2} of the main text. Thus, the expected form of the finite-size corrections is
\begin{equation}
\label{eq:app:SL-corrections}
S(L) / \ell_0 = a_1 (L / \ell_0)^{1/3}-a_2,\qquad L \gg \ell_0,
\end{equation}
with $a_{1,2} = O(1)$. This is Eq.~\eqref{eq:S-scaling} of the main text, with the leading correction additionally included.

Another source of finite-$L$ corrections is the quantum corrections, which were discussed in Sec.~\ref{sec:QuantumCorrections} above. Our analysis there indicates that quantum corrections lead to a correction to the superdiffusive asymptotics of the same form as in Eq.~\eqref{eq:app:SL-corrections} but of opposite sign. The numerical results show (see the analysis below) that, for finite $\gamma$ (class AIII), the total correction is dominated by ballistic effects described by Eq.~\eqref{eq:app:SL-corrections} with $a_2 > 0$. On the other hand, the quantum corrections become more important for the $\gamma = \infty$ point, where the crossover to the BDI symmetry class (with a larger magnitude of quantum corrections) occurs.

\begin{figure}
    \centering
    \includegraphics[width=0.8\columnwidth]{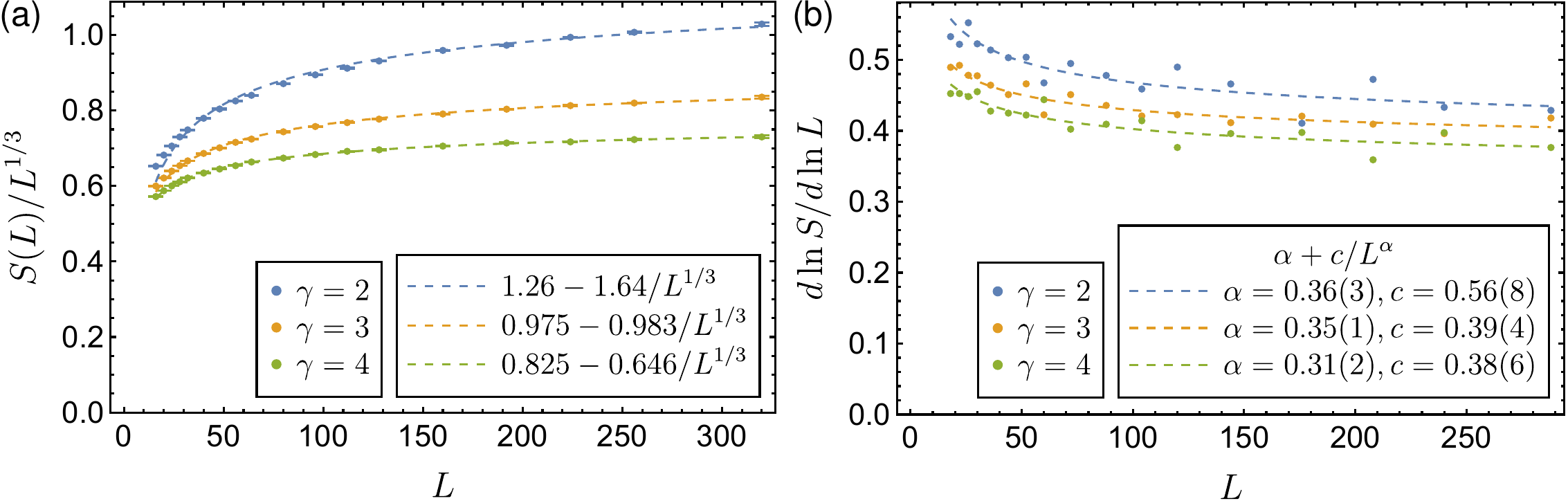}
    \caption{Numerical analysis of finite-size corrections to the asymptotic superdiffusive scaling of the half-system entanglement entropy $S(L)$ for measurement rates $\gamma = 2,3,4$. These plots use the same data as shown in Fig.~\ref{fig:ent_collapse} of the main text. (a)~Ratio $S(L) / L^{1/3}$ as a function of $L$; dashed lines: best fits according to Eq.~\eqref{eq:app:SL-corrections}. (b)~Logarithmic derivative $d \ln S / d \ln L$ as a function of $L$; dashed lines: best fits according to Eq.~\eqref{eq:app:SL-corrections-fit}. The extrapolated values of the power-law exponent $\alpha$ obtained from the fits are in very good agreement with the analytically predicted value $\alpha = 1/3$.}
    \label{fig:corrections}
\end{figure}

We now turn to the analysis of the numerical data. We focus on sufficiently large values of the measurement rate, $\gamma = 2, 3, 4$, for which the values of the ratio $L / \ell_0$ are large enough, so that the system is reasonably close to the superdiffusive limit [$d \ln S / d \ln L < 0.55$, see Fig.~\ref{fig:ent_collapse}(b)] in the range of considered system sizes $L$. For small $\gamma$, the ratio $L / \ell_0$ is not large enough and the system is too far from the asymptotic superdiffusive limit for accessible system sizes, so that Eq.~\eqref{eq:app:SL-corrections} describing the approach to this limit cannot be used.

In Fig.~\ref{fig:corrections}(a), we plot the ratio $S / L^{1/3}$, which is analytically predicted to saturate at a constant value for $L \to \infty$, as a function of $L$. We observe a nearly perfect agreement with the expected behavior \eqref{eq:app:SL-corrections} that involves the leading finite-size correction, in the whole range of available system sizes.
It is worth noting that for small system sizes, the correction is rather large: it reaches  
$\sim 50\%$ for $\gamma = 2$,
$\sim 40\%$ for $\gamma = 3$, and
$\sim 30\%$ for $\gamma = 4$.
Our analysis shows that the deviations from the superdiffusive asymptotic  ($L / \ell_0\to \infty$) behavior observed in Fig.~\ref{fig:ent_collapse} can be perfectly explained by analytically predicted finite-size corrections.

As a complementary approach to the analysis of finite-size corrections in numerical data, we inspect in Fig.~\ref{fig:corrections}(b) the logarithmic derivative $d\ln S / d\ln L$. 
We assume the scaling form equivalent to Eq.~\eqref{eq:app:SL-corrections} but replacing the exponent 1/3 by a free parameter $\alpha$,
\begin{equation}
\label{eq:app:SL-corrections-fit-1}
S(L)=s\left(L^{\alpha}-b\right),\qquad L \gg \ell_0 \,.
\end{equation}
This yields for the logarithmic derivative
\begin{equation}
\label{eq:app:SL-corrections-fit}
\frac{d\ln S}{d\ln L}=\alpha+\alpha b/L^{\alpha} \equiv \alpha+c/L^{\alpha}, \qquad L \gg \ell_0 \,.
\end{equation}
We fit numerical data for the logarithmic derivative to Eq.~\eqref{eq:app:SL-corrections-fit}, keeping $\alpha$ and $c$ as fitting parameters. Even though the numerical differentiation substantially magnifies the statistical noise in the data, we observe a reasonably good quality of fits. The extracted values of the power-law exponent $\alpha = 0.31(2), 0.35(1), 0.36(3)$ are perfectly consistent with the predicted value $\alpha = 1/3$.

Summarizing, our analysis of finite-size effects in numerical data for the entanglement entropy provides an additional strong support of the analytically predicted superdiffusive asymptotic behavior of $S(L)$ with the exponent 1/3 and for the form \eqref{eq:app:SL-corrections} of the leading finite-size correction. 

As a final remark, we note that a slow decay 
$\sim (\ell_0 / L)^{1/3}$ of the relative finite-size correction in Eq.~\eqref{eq:app:SL-corrections} implies that very large system sizes would be needed to approach the asymptotics of $S(L)$ with a given high accuracy. As an example, for the data shown in Fig.~\ref{fig:corrections} (i.e., for sufficiently large $\gamma$), system sizes needed to reach the asymptotics with 5\% accuracy can be estimated as $L \sim 10^4$, and the 1\% accuracy will be reached only for $L \sim 10^6$ (which is clearly not accessible numerically). For smaller values of $\gamma$ (i.e., larger $\ell_0$), the required system sizes are even larger.

\end{document}